\documentclass{JFM-FLM_Au_arxiv}
\makeatletter
\let\For\relax
\let\EndFor\relax
\let\State\relax

\let\Require\relax

\makeatother
\usepackage{algorithmicx}
\usepackage{algpseudocode}
\usepackage{bm}
\usepackage{hyperref}
\usepackage{comment}
\usepackage{tabularx}
\usepackage{graphicx}

\usepackage{multirow}

\lefttitle{Raimondi et al.}
\righttitle{Cilia-driven transport in confined ducts}

\newcounter{algorithm}
\renewcommand{\thealgorithm}{B.\arabic{algorithm}}

\newcommand{\algcaption}[2]{%
  \refstepcounter{algorithm}%
  \vspace{0.5em}%
  \noindent{Algorithm \thealgorithm.} #2\par
  \label{#1}%
  \vspace{0.5em}%
}

\title{Flow Transport in Active Porous Media}
\title{Transport in active porous media: a continuum model for cilia-driven flows}
\title{Cilia-driven fluid transport in confined ducts}
\title{Cilia-driven transport in confined ducts: an active porous media model}

\author{JP Raimondi\aff{1}, Feng Ling\aff{1,2}, \and Eva Kanso \aff{1,3} }

\affiliation{\aff{1} Department of Aerospace and Mechanical Engineering,  \\ University of Southern California, Los Angeles, California 90089, USA
\aff{2} School of Physics, Nankai University, Nankai District, Tianjin 300071, China \\
\aff{3} Department of Physics and Astronomy,  \\ University of Southern California, Los Angeles, California 90089, USA
}

\corresau{Eva Kanso, kanso@usc.edu}

\begin{document}
\maketitle

\begin{abstract}
Ciliated organs transport viscous fluids through confined ducts, yet how duct morphology and ciliary activity jointly set the limits of flow rate and sustainable pressure remains unclear. Here, we model dense arrays of beating cilia lining duct walls as an active porous medium driven by prescribed metachronal waves, and identify two key morphological parameters that govern transport: the ciliary confinement ratio and the mean ciliary fraction. The resulting flows are described by the incompressible Navier--Stokes--Brinkman equations, which we solve numerically using a spectral method in the low-Reynolds-number regime. We also develop a complementary mean-field analytical model. The active porous medium framework provides an intermediate description between classical envelope theories and filament-resolved simulations and enables a systematic investigation of how fluid transport is shaped by confinement and packing of ciliary material. We find that transport is characterized by a decreasing linear relationship between flow rate and pressure generation, marking a fundamental trade-off between throughput and sustainable adverse pressure. These results provide a unified physical interpretation of the morphological diversity of ciliated ducts, from high-throughput ciliary carpets to pressure-generating ciliary flames, and offer guiding principles for the design of bio-inspired microfluidic pumps.
\end{abstract}

\begin{keywords} ciliary flows, active porous media, low-Reynolds-number flows, biological transport, Brinkman equations
\end{keywords}

\clearpage


\section{Introduction}
\label{sec:intro}

Ciliated organs drive essential transport processes throughout animal physiology. In humans, coordinated ciliary beating directs cerebrospinal fluid in the brain ventricles~\citep{faubel2016}, transports gametes in the reproductive tract~\citep{greenstone1985,satir2007overview}, and clears mucus in the airways~\citep{afzelius1976,roth2025structure}. These functions arise from the collective activity of dense arrays of motile cilia. Motile cilia are micron-scale active filaments whose oscillatory motion generates fluid flow in the viscous regime at vanishingly small Reynolds numbers~\citep{Lauga2009,gilpin2020multiscale}. Despite decades of work on ciliary beat kinematics and synchronization, a fundamental question remains open: how do ciliary activity and duct morphology jointly set the limits of fluid transport, in terms of achievable flow rate and pressure generation? Recent evidence points to a tight coupling between ciliated duct morphology and transport function across diverse biological systems~\citep{ling2024flow}, yet a predictive quantitative framework linking beat kinematics, metachronal wave coordination, and duct geometry to fluid transport remains incomplete.

\begin{figure}
    \centering
    \includegraphics[width=\linewidth]{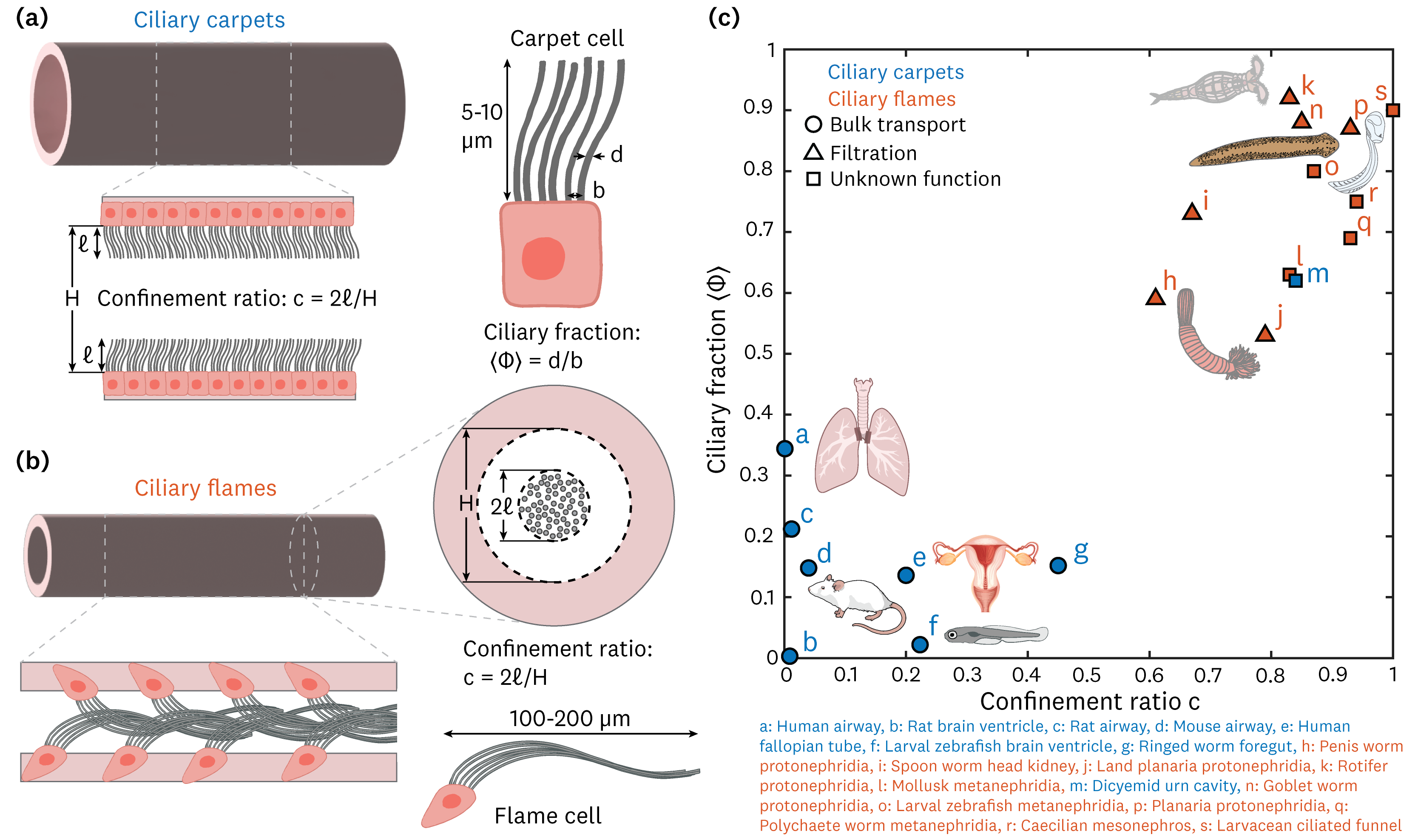} 
    \caption{\textbf{Ciliated systems, corresponding geometric measures, and biological data.} The ciliary confinement ratio, $c$, denotes the ratio of the ciliary layer thickness to the lumen diameter. The ciliary fraction, $\langle \Phi \rangle$, defines the corresponding fraction of cilia within the ciliated region.
(a) \textbf{Ciliary carpets} tend to have large lumen ($50$–$1000,\mu\mathrm{m}$) with short ($5$–$10,\mu\mathrm{m}$), wall-normal cilia. Activity is distributed over the boundary and quantified from cross-sections parallel to the lumen axis.
(b) \textbf{Ciliary flames} have small lumen ($1$–$50,\mu\mathrm{m}$) with long ($100$–$200,\mu\mathrm{m}$), axially aligned cilia. Ciliary activity is concentrated near the center of the lumen and quantified from cross-sections perpendicular to the axis.
(c) Measurements of $c$ and $\langle \Phi \rangle$ from biological data, showing distinct regimes: carpets occupy low $c$ and $\langle \Phi \rangle$, while flames exhibit high values of both. Marker shape indicates function: circles for bulk transport, triangles for filtration, and squares for ducts with unknown functions. See table \ref{tab:biological data} for full dataset and references. 
Lung and rat illustrations from NIAID NIH BioArt Source (bioart.niaid.nih.gov/bioart/231) and (bioart.niaid.nih.gov/bioart/54), female reproductive tract \copyright~blueringmedia / Adobe Stock, rotifer \copyright~Kazakova Maryia / Adobe Stock, others are original. 
}
    \label{fig:intro}
\end{figure}

In vertebrate organs, short cilia form dense carpets that beat predominantly normal to the epithelial surface, generating metachronal waves well suited for high-throughput transport and mixing in wide lumens~\citep{Ding2014, Nawroth2017, gilpin2020multiscale, kanale2022spontaneous} (figure~\ref{fig:intro}a). In contrast, some invertebrates exhibit ciliary flames, in which long, tightly packed cilia beat longitudinally within narrow ducts, pumping fluid through highly resistive pathways in filtration and excretion organs~\citep{vogel2007living,ling2024flow} (figure~\ref{fig:intro}b). Although these configurations differ markedly in geometry and function, both operate in the viscous regime and rely on collective traveling-wave kinematics to drive transport. Comparative surveys across animal phyla indicate that ciliary carpets and ciliary flames represent two extremes in ciliary architecture~\citep{ling2024flow}: carpets are adapted for high-throughput transport and mixing, whereas flames are adapted for filtration and sustained pumping against adverse pressure gradients. Together, these systems illustrate how duct geometry and ciliary organization map onto distinct transport tasks, motivating the need for a unified physical description capable of spanning this diversity.

To place these architectures on a quantitative footing, we compile geometric data from published measurements, drawing in particular on \cite{ling2024flow}. We characterize each system using two dimensionless parameters: the ciliary confinement ratio $c$, defined as the ratio of the ciliary layer height to the total duct height, and the mean ciliary fraction $\langle \Phi \rangle$, defined as the fraction of the ciliary layer occupied by ciliary material. The compiled data (figure~\ref{fig:intro}c, table~\ref{tab:biological data}, and appendix~\ref{App:biodata}) show that ciliary carpets occupy a regime of low confinement and moderate ciliary fraction, whereas ciliary flames lie at high confinement and high ciliary fraction. 

These morphological trends 
raise a natural question: given a ciliated channel characterized by confinement ratio $c$ and mean ciliary fraction $\langle \Phi \rangle$, how do these geometric parameters, together with prescribed ciliary kinematics, determine the resulting flow rate and the maximum pressure that can be sustained?

Classical theoretical approaches to cilia-driven flows trace back to Taylor’s swimming sheet~\citep{Taylor1951} and Blake’s envelope model~\citep{Blake1971,Blake1972}, in which densely packed ciliary carpets are represented as an impermeable surface supporting traveling waves. These models, and their extensions~\citep{Michelin2011,Smith2008,Chrispell2013,Ramirez2020,liu2025flow, liu2025optimal, liu2025nutrient, liu2026feeding}, have provided fundamental insight into propulsion and transport, but neglect flow penetration through the ciliary layer and its internal structure. 
At the opposite extreme, filament-resolved models capture individual cilium mechanics and hydrodynamic interactions~\citep{Liron1976,Liron1978,Gueron1992,Blake1982,Ding2014,Guo2014,Nawroth2017,Guo2020,osterman2011finding}, but become computationally prohibitive and parameter-rich when applied to realistic systems.

A promising intermediate description models dense ciliary assemblies as an active porous medium governed by a Brinkman-type equation~\citep{Stein2019, wuttanachamsri2020effects, wuttanachamsri2021effects, ling2024flow, dutta2024self}. In this framework, the collective effect of many beating filaments is coarse-grained into a distributed body force and an effective drag.

Here, we build on this approach to develop a unified Navier–Stokes–Brinkman model for ciliary pumping in confined channels. The channel is represented as a multi-layer system, with ciliary regions modeled as active porous layers endowed with prescribed traveling-wave kinematics. This formulation occupies an intermediate level between envelope models and filament-resolved simulations, retaining permeability and distributed forcing while avoiding explicit resolution of individual cilia. It enables a systematic investigation of how cilia length, packing density, metachronal coordination, and channel geometry jointly determine transport across the full spectrum from carpets to flames.

\begin{figure}
    \centering
    \includegraphics[scale=1]{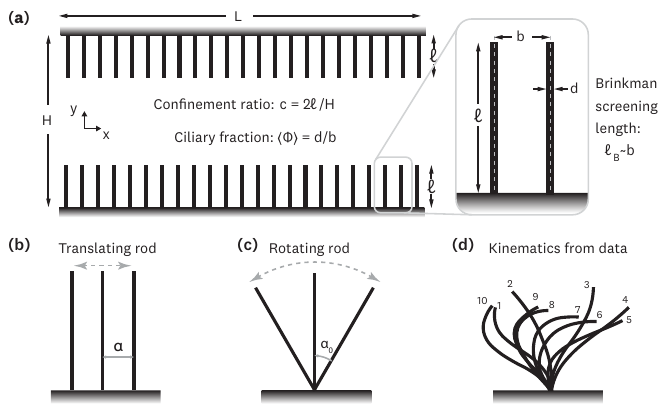}
    \caption{\textbf{Reference configuration and ciliary kinematics.} 
(a) Undeformed reference configuration of the active Brinkman channel, with channel length $L$, height $H$, and ciliary layer thickness $\ell$. The ciliary confinement ratio is defined as $c = 2\ell/H$. Inset: cilia of diameter $d$ are uniformly spaced with center-to-center distance $b$, giving an average ciliary fraction $\langle \Phi \rangle = d/b$. The Brinkman screening length $\ell_B$ is proportional to $b$. 
(b) Translating-rod kinematic model with oscillation amplitude $a$. 
(c) Rotating-rod kinematic model with oscillation amplitude $\alpha_0$. 
(d) Experimentally measured ciliary kinematics in rabbit trachea~\citep{sanderson1981ciliary}; snapshots of the motion are evenly spaced over one beat cycle.
}
    \label{fig:ref config and kin}
\end{figure}



\section{Problem formulation}

We consider cilia-driven flow in a rectangular channel of length $L$ and height $H$, with ciliary carpets lining the side walls (figure~\ref{fig:ref config and kin}). Lengths are nondimensionalized by $L$ and time by $L/U$, where $U=\omega \ell$ is a characteristic ciliary speed defined by the beat frequency $\omega$ and ciliary layer thickness $\ell$.

\subsection{Fluid model}

We adopt a coarse-grained continuum description in which the coupled ciliary activity and fluid flow are modeled as an active porous medium~\citep{Stein2019}. The dynamics are governed by the incompressible Navier–Stokes equations with an active Brinkman drag term, given by the nondimensional form
\begin{equation} \label{NSB}
\begin{split}
\mathrm{Re}\left(\frac{\partial {\bm{u}}}{\partial t}
+ {\bm{u}}\cdot\nabla{\bm{u}}\right)
& = -\nabla p
+ \nabla^{2}{\bm{u}}
- B\bigl({\bm{u}}-{\bm{v}}_{c}\bigr)
- \Delta P\,\bm{e}_{x},\qquad
\nabla \cdot {\bm{u}} = 0.
\end{split}
\end{equation}
Here, $t$ denotes time, and $x$ and $y$ are the streamwise and transverse coordinates, respectively, with $\bm{e}_x$ the unit vector in the streamwise direction. The fluid velocity and pressure fields are denoted by $\bm{u}(x,y,t)$ and $p(x,y,t)$, with the latter enforcing incompressibility.

The Reynolds number is defined as $\mathrm{Re} = \rho U L / \mu$, where $\rho$ and $\mu$ are the fluid density and viscosity. A uniform pressure gradient $\Delta P$ is imposed opposite to the desired direction of transport.

In~\eqref{NSB}, the term $B(\bm{u}-\bm{v}_c)$ represents an effective momentum exchange between the fluid and the actively driven porous medium. The contribution $-B\bm{u}$ corresponds to viscous drag passively induced by the porous medium, while $B\bm{v}_c$ acts as an internal forcing that can drive flow even in the absence of external pressure gradients. Together, these terms act to force the fluid velocity $\bm{u}$ to match the imposed porous medium velocity $\bm{v}_c$ with the strength of this forcing being set by the Brinkman coefficient, $B$. Both $\bm{v}_c(x,y,t)$ and $B(x,y,t)$ vary in space and time as prescribed by the ciliary motion (section~\ref{sec:cilia}) and are nonzero only within the ciliary layers, vanishing elsewhere.

The field $\bm{v}_c$ encodes the ciliary kinematics, including individual beating and metachronal wave propagation. These motions deform the porous medium and thereby modulate the local ciliary fraction $\Phi(x,y,t)$, defined as the fraction of the ciliated layer occupied by ciliary material. The Brinkman coefficient $B$ depends on $\Phi$ and quantifies viscous drag.

In the Stokes–Brinkman limit ($\mathrm{Re}=0$), this drag replaces the long-ranged Stokeslet Green’s function of the Stokes equations, which decays as $1/r$, with a screened Green’s function that decays as $e^{-r/\ell_B}/r$. The parameter $\ell_B$ is the Brinkman hydrodynamic screening length, which sets the distance over which flow disturbances propagate in the effective porous medium. In the present model, $\ell_B = \langle B \rangle^{-1/2}$, where $\langle B \rangle$ denotes the spatial average of the Brinkman coefficient.

To close the model, we prescribe planar ciliary motion (section~\ref{sec:cilia}) and consider periodic boundary conditions in the longitudinal $x$-direction, no-slip boundary conditions in the transverse $y$-direction, and assume that the flow is invariant in the out of plane $z$-direction, such that
\begin{equation} 
\begin{split}
	\bm{u}(0,y,t) & =  \bm{u}(L,y,t), \quad
	\bm{u}(x,0,t) = \bm{u}(x,H,t) = \bm{0}, \quad p(0,y,t) = p(L,y,t).
    \label{eq:bc}
\end{split}
\end{equation}

Cilia-driven flows typically occur at low Reynolds numbers, $\mathrm{Re}=10^{-4}\text{--}10^{-1}$, where viscous forces dominate and inertial effects are weak~\citep{Ding2014, Nawroth2017, gilpin2020multiscale, kanale2022spontaneous}. However, unsteady beating and strong confinement can generate finite advective and inertial contributions~\citep{Wei2019, wei2021measurements}. 

From a numerical standpoint, solving the time-dependent Navier–Stokes equations at small but finite Reynolds number provides a stable and well-conditioned framework for capturing unsteady low-Re flows. We therefore retain the full Navier–Stokes–Brinkman system~\eqref{NSB} and fix $\mathrm{Re}=0.01$, which remains in the viscous-dominated regime while allowing for weak inertial effects. Here, we use an implicit-explicit spectral method to solve~(\ref{NSB},\ref{eq:bc}). Details of the numerical method, together with convergence tests and validation, are provided in appendices~\ref{App:numerics} and~\ref{App:validate}, algorithms~\ref{alg:precompute} and~\ref{alg:flow}, and figures~\ref{fig:convergence testing}--\ref{fig:k flow validation}.

\begin{table}
\captionsetup{skip=0pt}
\caption{\textbf{Ciliated channel parameters:} Representative dimensional values compiled from the literature~\citep{gilpin2020multiscale, ling2024flow, roth2025structure}, together with their nondimensional counterparts obtained using the characteristic channel length $L$, ciliary speed $U=\omega \ell = 100\,\mu\mathrm{m\,s^{-1}}$, and fluid viscosity $\mu = 10^{-3}\,\mathrm{N\,s\,m^{-2}}$. Average values of the ciliary fraction and Brinkman coefficient, denoted by $\langle \cdot \rangle$ and based on biological observations, are used in the reference configuration.}
\label{tab:param}
\centering
\setlength{\tabcolsep}{5pt}
\renewcommand{\arraystretch}{1.2}
\begin{tabular}{@{}l|l|ll|l@{}}
\cline{1-5}
{\textbf{Quantity}} & {\textbf{Symbol}} & \multicolumn{2}{c|}{\textbf{Dimensional value}}  & \textbf{Non-dimensional value} \\  
\cline{1-5}
Channel length & $L$ & 100 & $\mu{\rm m}$ & 1\\
Channel height & $H$ & 20-400 & $\mu{\rm m}$ & 0.2-4\\
Ciliary layer thickness & $\ell$ & 10 & $\mu{\rm m}$  & \\
Ciliary confinement ratio & $c = 2\ell/H$ & \multicolumn{2}{c|}{------} & 0-1\\
Ciliary fraction & $ \Phi(x,y,t)$ &  \multicolumn{2}{c|}{------} & $\langle \Phi \rangle=$ 0-1\\ 
Brinkman coefficient & $ B(x,y,t) $ & $\langle B \rangle =10^7-10^{11}$ & ${\rm N}\cdot {\rm s}\cdot {\rm m}^{-4}$  & $\bar{B} = \langle B \rangle =$ 10$^2$-10$^6$\\
Adverse pressure & $\Delta P$ & 0-400 & ${\rm N} \cdot {\rm m}^{-2}$  & 0-10$^5$\\
\end{tabular}
\end{table}

\begin{table}
\captionsetup{skip=0pt}
\caption{\textbf{Scaling from nondimensional to dimensional values.} Lengths are scaled by the channel length $L$, velocities by $U=\omega \ell$, time by $L/U$, pressure by $\mu U/L$, and the Brinkman coefficient by $\mu/L^2$.
Throughout, dimensional values are obtained using $L = 100 \ \mu{\rm m}$, $U = 100\,\mu\mathrm{m\,s^{-1}}$, and  $\mu = 10^{-3}\,\mathrm{N\,s\,m^{-2}}$.}
\label{tab:nondim_param}
\centering
\begin{tabular}{@{}lll@{}}
\cline{1-3}
\textbf{Quantity} & \textbf{Non-dimensional} & \textbf{Scaling to dimensional}\\
\cline{1-3}
Streamwise Eulerian coordinate & $x$ & $Lx$ \\
Wall-normal Eulerian coordinate & $y$  & $ Ly$ \\
Time & $t$ & $(L/U)t$ \\
Fluid velocity vector field & $\bm{u}(x,y,t)$ & $U \bm{u}$ \\
Fluid pressure field & $p(x,y,t)$ & $(\mu U/L) p$ \\
Ciliary velocity vector field & $\bm{v}_c(x,y,t)$ &  $U\bm{v}_c$ \\
Brinkman drag coefficient &  $B(x,y,t)$ & $(\mu/L^2)B$ \\
Adverse pressure gradient & $\Delta P$ &  $ (\mu U/L)\Delta P$ \\
\end{tabular}
\end{table}

\subsection{Ciliary kinematics}
\label{sec:cilia}

We next outline a procedure for constructing the continuum fields in~\eqref{NSB}: the Brinkman coefficient $B(x,y,t)$ and the ciliary velocity $\bm{v}_c(x,y,t)$, both determined by the geometry and kinematics of the ciliary system.

It is convenient to introduce an undeformed, uniform reference configuration consisting of a discrete ciliary carpet extending over the channel length $L$, with ciliary layer thickness $\ell$, diameter $d$, and intercilium spacing $b$. In this uniform state, the ciliary fraction is
\begin{equation}
\langle \Phi \rangle = \frac{L}{b}\frac{A_c}{A} = \frac{d}{b},
\label{eq:phi_ref}
\end{equation}
where $L/b$ is the number of cilia, $A_c = \ell d$ is the area of a single cilium, and $A = \ell L$ is the total carpet area.

To set the scale of the mean Brinkman drag coefficient $\langle B \rangle$, we estimate the drag force coefficient in the ciliary layer as follows. Using slender-body theory, we approximate the drag force  $F_D = c_D U$ on a single cilium in a uniform flow of speed $U$ directed perpendicular to its axis. The drag coefficient is $c_D = {4 \pi \mu \ell}/{\ln(2 \ell /d)}$ (with $\mu=1$ in the nondimensional formulation). Neglecting hydrodynamic interactions with the anchoring wall and between neighboring cilia, the total drag exerted by the ciliary layer scales linearly with the number of cilia. The corresponding drag coefficient is therefore $C_{D} = c_D L/b={4 \pi \mu \ell L}/{b\ln(2 \ell /d)}$.

The drag coefficient $C_D$ converts velocity to force 
while the Brinkman coefficient $B$ converts velocity to pressure gradient. 
Thus, to obtain the Brinkman coefficient $\langle B \rangle$, we divide $C_D$ by an effective volume. Given that drag acts on the pore scale $(b-d)$ and is distributed over the carpet area $\ell L$, we divide $C_D$ by $\ell L (b-d)$. Substituting~\eqref{eq:phi_ref} into the expression for $C_D$, we arrive at the Brinkman drag coefficient in the undeformed reference configuration 
\begin{equation}
\langle B \rangle
= \frac{4 \pi \mu} {d(b-d)
\ln(2 \ell/d)}\langle \Phi \rangle.
\end{equation}

Ciliary motion deforms the reference configuration, giving rise to spatially and temporally varying fields $B(x,y,t)$ and $\bm{v}_c(x,y,t)$ that reflect both the geometry and kinematics of the ciliary carpet. We therefore seek expressions relating these Eulerian fields to the reference Brinkman coefficient $\langle B \rangle$ and the prescribed ciliary motion. 

Let $(X,Y)$ denote Lagrangian material coordinates, and let the reference and current (deformed) configurations of the cilia be related by a forward kinematic mapping $\bm{\chi}_c(X,Y,t)$ that maps each material point $(X,Y)$ to its Eulerian position $(x,y)$ at time $t$. Conversely, the inverse mapping $\bm{\chi}_c^{-1}(x,y,t)$ assigns each fixed Eulerian point $(x,y)$ to the corresponding Lagrangian label $(X,Y)$ of the cilium occupying that location at time $t$ 
\begin{equation} \label{inverse mapping}
\bm{\chi}_c(X,Y,t) 
= \begin{bmatrix} x(X,Y,t) \\[1ex] y(X,Y,t) \end{bmatrix}, \qquad
\bm{\chi}_c^{-1}(x,y,t) 
= \begin{bmatrix} X(x,y,t) \\[1ex] Y(x,y,t) \end{bmatrix}.
\end{equation}
The ciliary velocity field in the Lagrangian frame is given by
\begin{equation}
\bm{v}_{c}(X,Y,t) = \frac{\partial \bm{\chi}_c(X,Y,t)}{\partial t}.
\label{eq:Vc}
\end{equation}
The deformation gradient is $\bm{F}=\nabla_{\bm{X}}\bm{\chi}_c$ with determinant $J=\det(\bm{F})$, which measures the local area change induced by the ciliary deformation. This allows the ciliary fraction $\langle \Phi \rangle$ and Brinkman coefficient to be mapped from the reference to the deformed configuration, yielding
\begin{equation} \label{phi push-forward}
B(X,Y,t) = \frac{\langle B\rangle}{J(X,Y,t)}.
\end{equation}
In regions where the active porous medium is locally compressed ($J<1$), the Brinkman drag $B$ is enhanced relative to the reference state, whereas in stretched regions ($J>1$) it is reduced.

In~\eqref{eq:Vc} and~\eqref{phi push-forward}, the  fields $\bm{v}_{c}$ and $B$ are expressed in terms of Lagrangian coordinates $(X,Y)$, while the Navier-Stokes-Brinkman equations~\eqref{NSB} and boundary conditions~\eqref{eq:bc} are formulated in Eulerian coordinates $(x,y)$. Using the inverse mapping $\bm{\chi}_c^{-1}(x,y,t)$, we obtain
\begin{equation} \label{current Phi}
\bm{v}_c(x,y,t) 
= \left. \frac{\partial \bm{\chi}_c}{\partial t}\right|_{(X(x,y,t),\,Y(x,y,t))}, 
\qquad
B(x,y,t) 
= \left. \frac{\langle B \rangle}{J}\right|_{(X(x,y,t),\,Y(x,y,t))}.
\end{equation}
In summary, this procedure defines a uniform reference configuration and a kinematic mapping $\boldsymbol{\chi}_c$ that together yield continuous fields $B(x,y,t)$ and $\bm{v}_c(x,y,t)$ capturing the spatiotemporal variations induced by ciliary motion. The framework is general and can be extended to three-dimensional settings and heterogeneous reference configurations. We apply it in section~\ref{sec:simulations} to idealized kinematic models and experimentally reconstructed ciliary motion.

Lastly, for later use, we introduce a measure of the total ciliary material in the channel defined by the spatial integral of the local ciliary fraction $\Phi(x,y,t)$. Conservation of material implies that the total ciliary content is invariant under deformation. Accordingly, the integral of $\Phi$ over the current configuration is equal to that of the reference configuration, yielding a total ciliary material proportional to $2\ell L \langle\Phi\rangle$. For fixed cilia length $\ell$ and channel length $L$, the reference ciliary fraction $\langle \Phi \rangle$ therefore uniquely determines the total amount of ciliary material in the system.

\section{Numerical simulations}
\label{sec:simulations}

To demonstrate the versatility of our modeling framework, we consider three planar models of ciliary motion with increasing kinematic complexity (figure~\ref{fig:ref config and kin}b-d and table~\ref{tab:kinematic_models}). In all cases, individual cilia undergo periodic oscillations at a frequency $\omega$ with a systematic phase lag along the longitudinal $\bm{e}_x$-direction of the carpet, giving rise to a traveling metachronal wave. The metachronal wavenumber $k$ sets the phase lag between neighboring cilia and thus the wavelength of the traveling wave. Throughout, we take the wavenumber $k = {2\pi}/{L}$,
so that one metachronal wavelength spans the channel length $L$, with positive $k$ corresponding to waves propagating from right (positive $x$) to left (negative $x$). Finally, we restrict the kinematic parameters such that the mapping $\bm{\chi}$ remains locally invertible in all cases and the active porous medium does not fold over itself; that is, we constrain the Jacobian to be strictly positive $J>0$.



\begin{figure} [t]
    \centering
    \includegraphics[width=\linewidth]{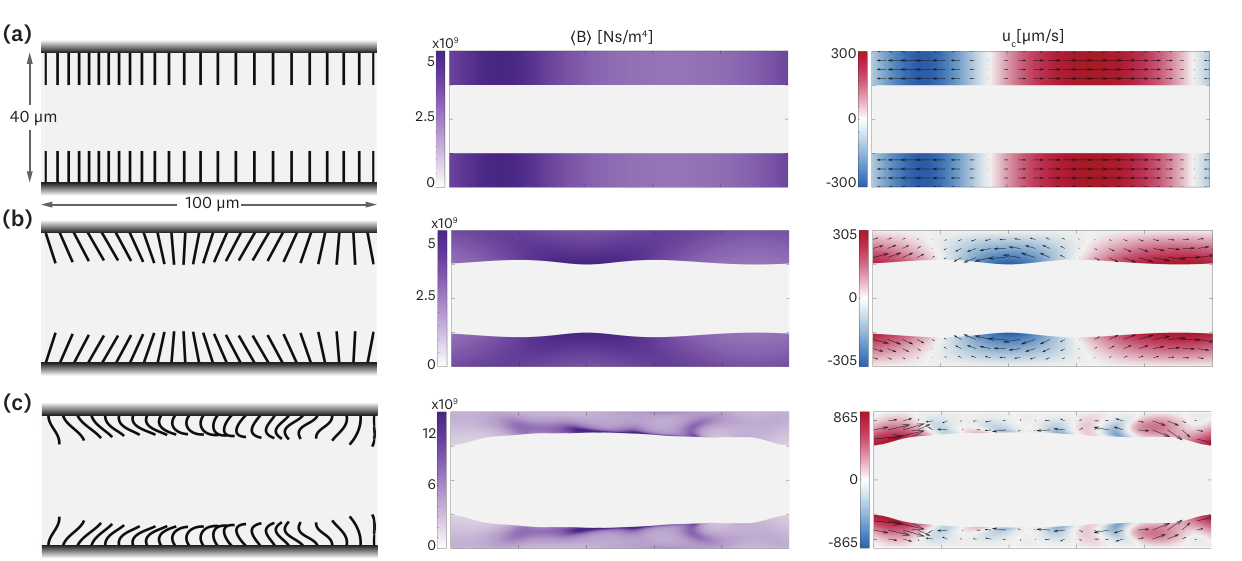}
    \caption{\textbf{Discrete ciliary carpets and corresponding continuum fields.} Left to right: time snapshots of the discrete rod representation (number of rods not to scale), Brinkman coefficient field, and ciliary velocity field (colormap showing streamwise component $u_c = \bm{v}_c\cdot\mathbf{e}_x$). (a) Translating rod model, with $a = 5 \rm \,\mu m$. (b) Rotating rod model, with $\alpha_0 = \pi/6$. (c) Kinematics reconstructed from experimental measurements~\citep{sanderson1981ciliary}.
In all cases, a single metachronal wave $\theta(X,t)=\omega t+kX$ with $k=2\pi/L$ is imposed over the channel length. Parameter values: $\omega = 10 \ \rm{Hz},$ $b = 1 \rm \,\mu m$, $d = 0.2 \rm \,\mu m$, $\ell = 10 \rm \, \mu m$, $L = 100 \rm \, \mu m$, $H = 40 \rm \, \mu m$,  $\langle \Phi \rangle = 0.2$, and $\langle B\rangle =3.4 \times 10^5$.}
    \label{fig:ciliakinematics}
\end{figure}

The simplest model (figure~\ref{fig:ref config and kin}b) represents each cilium as a rigid vertical rod undergoing symmetric horizontal oscillations, with a spatially varying phase $\theta(X,t)$ corresponding to a metachronal wave of wavenumber $k$ in the longitudinal direction,
\begin{equation} \label{translating rod}
\bm{\chi}_c(X,Y,t) = \begin{bmatrix} X + a \cos\theta \\ Y \end{bmatrix}, \qquad  \theta(X,t)=\omega t+kX.
\end{equation}
Here, $X$ labels the position along the carpet and thus indexes individual cilia, while $Y$ denotes arclength along each cilium measured from its base. The parameters $a$ and $\omega$ are the non-dimensional oscillation amplitude and frequency, respectively.  The  condition $J>0$ reduces to $ak<1$. Because the mapping \eqref{translating rod} is transcendental in $X$, its inverse cannot be obtained in closed form. We therefore compute $\bm{\chi}_c^{-1}$ numerically using a pointwise one-dimensional Newton iteration and evaluate the Eulerian fields $\bm{v}_c(x,y,t)$ and $B(x,y,t)$ (algorithm~\ref{alg:precompute}). Figure~\ref{fig:ciliakinematics}a depicts snapshots of the Brinkman drag field (center column) and corresponding ciliary velocity field (right column). Both fields translate smoothly and periodically in the direction of metachronal wave propagation, from right to left. The Brinkman field captures the local stretching and compression of the carpet: compressed regions correspond to cilia in their backward stroke, whereas stretched regions correspond to cilia in their forward stroke.

\begin{table}
\captionsetup{skip=0pt}
\caption{\textbf{Cilia beat kinematics:} parameters defining the three planar kinematic models introduced in section~\ref{sec:cilia} and figures~\ref{fig:ref config and kin} and~\ref{fig:ciliakinematics}. In all cases, the phase lag $\theta(X,t)=\omega t+kX$, with $k=2\pi/L$, correspond to a single metachronal wavelength over the channel length $L$.}
\label{tab:kinematic_models}
\centering

\begin{tabular}{@{}lll@{}}
\cline{1-3}
\textbf{Model} & \textbf{Mapping} $\boldsymbol{\chi}_c$ & \textbf{Parameters}  \\ 
\cline{1-3}
Translation 
& $\big(X+a\cos\theta,\;Y\big)$ 
& $a,\;\omega,\; k$ 
\\
Rotation 
& $\big(X+Y\sin\alpha,\;Y\cos\alpha\big)$, $\quad \alpha=\alpha_0\sin\theta \qquad$ 
& $\alpha_0,\;\omega,\; k$ 
 \\
Kinematics from data 
& $\big(X+x_c(Y,\theta),\;y_c(Y,\theta)\big)$ 
& $A_{0x},A_{0y},A_x,B_x,A_y,B_y,\;\omega, \; k$ 
\end{tabular}
\end{table}

The second kinematic model describes a rigid cilium undergoing symmetric angular oscillations about its base (figure~\ref{fig:ref config and kin}c), also with a spatially varying phase $\theta(X,t)=\omega t+kX$,
\begin{equation} \label{rotating rod}
\begin{split} 
\bm{\chi}_c(X,Y,t) &= \begin{bmatrix} X + Y \sin \alpha \\ Y \cos \alpha\end{bmatrix}, \qquad 
\alpha = \alpha_o \sin\theta.
\end{split} 
\end{equation}
Here, $\alpha(X,t)$ is the instantaneous tilt angle, $\alpha_o$ is the angular amplitude, and $\omega$ and $k$ again set the beat frequency and metachronal wavelength. This motion results in carpet deformations in both the $x$ and $y$ directions (figure~\ref{fig:ciliakinematics}b), with no-slip boundary conditions at the channel walls.

The third and most realistic model (figure \ref{fig:ref config and kin}d) is based on experimentally measured planar ciliary waveforms, with asymmetric power and recovery strokes, and again with $\theta(X,t)=\omega t+kX$,
\begin{equation} \label{FB kinematics}
\begin{split}
\bm{\chi}_c(X,Y,t) &= \begin{bmatrix} X + x_c(Y,\theta)\\ y_c(Y,\theta) \end{bmatrix}.
\end{split}
\end{equation}
Here, the functions $x_c(Y,\theta)$ and $y_c(Y,\theta)$ are represented by polynomial expansions in arclength and Fourier series in phase $\theta(X,t)$, with coefficients $A_{0x}, A_{0y}, A_x, A_y, B_x, B_y$ optimized to fit experimental measurements of rabbit trachea ciliated epithelium~\citep{fulford1986muco, sanderson1981ciliary},
\begin{equation}
\begin{split}
x_c(Y,\theta) &=\;\ell\,\sum_{m=1}^{M}\left(\frac{Y}{\ell}\right)^{m}
\left[
\frac{1}{2}\,A_{0x}^{(m)}
\;+\;\sum_{n=1}^{N} \Big(
A_{x}^{(m,n)}\cos(n\theta)\;+\;B_{x}^{(m,n)}\sin(n\theta)
\Big)
\right], \\
y_c(Y,\theta)
&=\;\ell\,\sum_{m=1}^{M}\left(\frac{Y}{\ell}\right)^{m}
\left[
\frac{1}{2}\,A_{0y}^{(m)}
\;+\;\sum_{n=1}^{N}\Big(
A_{y}^{(m,n)}\cos(n\theta)\;+\;B_{y}^{(m,n)}\sin(n\theta)
\Big)
\right] .
\end{split}
\end{equation}
This representation captures the full time-dependent shape of each cilium and the propagation of metachronal waves along the carpet (figure \ref{fig:ciliakinematics}c). For the same reference ciliary fraction and metachronal wavelength, this ciliary kinematics yields a maximum Brinkman coefficient more than twice as large as that of the translating and rotating rod cases because it results in large ciliary fractions near the surface of the carpet. It also produces larger peak ciliary velocities, even with identical beat frequency and comparable oscillation amplitudes, owing to its distinct fast power strokes and slow recovery strokes.

With the inputs to the Navier-Stokes-Brinkman equation \eqref{NSB}, $B(x,y,t)$ and $\bm{v}_c(x,y,t)$, in hand, we solve \eqref{NSB} subject to boundary conditions \eqref{eq:bc} using an implicit-explicit spectral method that represents $\bm{u}$ with Fourier modes in the $x$ direction and sine modes in the $y$ direction, automatically satisfying \eqref{eq:bc}. The full details of our numerical method and validation are provided in appendices \ref{App:numerics}-\ref{App:validate}. Owing to the traveling wave form of the inputs, the flow field is also a traveling wave. Thus for each simulation, we advance the solution forward in time until a periodic steady state is reached.

\begin{figure} [ht!]
    \centering
    \includegraphics[width=\linewidth]{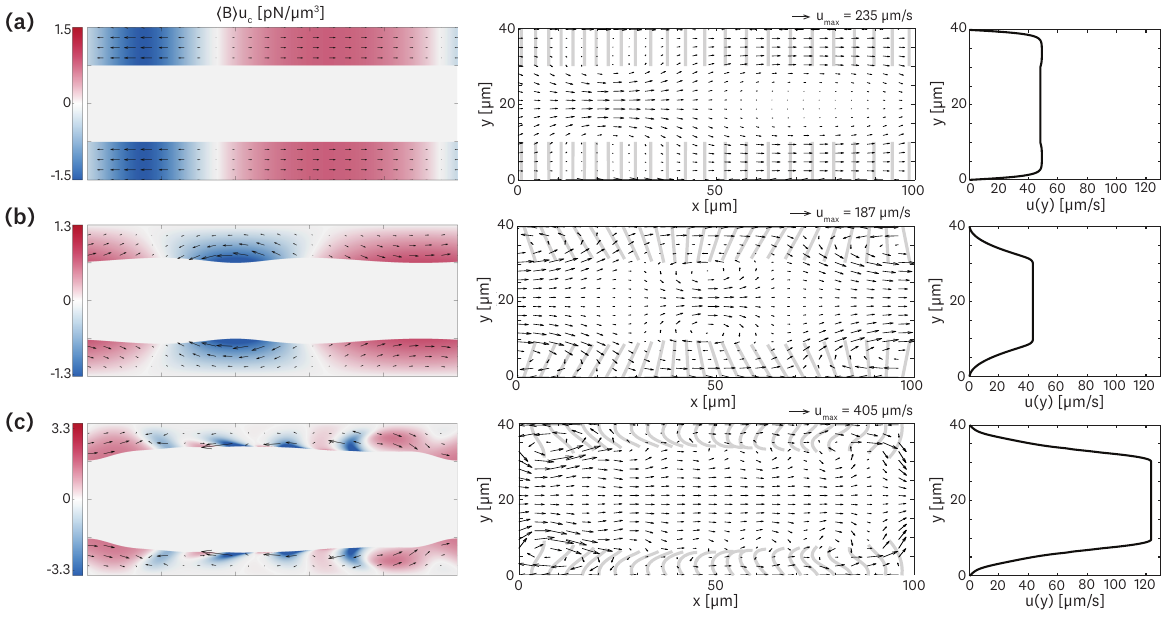}
    \caption{\textbf{Flows driven by prescribed ciliary activity.} Left: instantaneous active porous media force density, $B \bm{v}_c$, colormap indicates streamwise component $B \bm{v}_c \cdot \mathbf{e}_x$ in units of $\rm pN/\mu m^3$. Middle: instantaneous velocity fields at zero applied pressure, scaled by the maximum speed (indicated). Grey rods denote the extent of the ciliary carpet (rod spacing not to scale). Right: streamwise velocity profiles averaged over one beat cycle. Corresponding ciliary activity depicted in figure~\ref{fig:ciliakinematics} and table~\ref{tab:kinematic_models}. (a) Translating rod model resulting in pumping rate $Q = 1867 \, \mu\mathrm{m}^2\mathrm{s}^{-1}$. (b) Rotating rod model, $Q = 1228 \, \mu\mathrm{m}^2\mathrm{s}^{-1}$. (c) Experimentally reconstructed kinematics, $Q = 3540 \, \mu\mathrm{m}^2\mathrm{s}^{-1}$. $\Delta P = 0$, all other parameter values are the same as those in figure \ref{fig:ciliakinematics}.}
    \label{fig:flow fields}
\end{figure}

In figure~\ref{fig:flow fields}, we show snapshots of the active force density $B\bm{v}_c$ (left column) for the three ciliary kinematic models in figure~\ref{fig:ciliakinematics}, each with identical reference configurations and channel geometries. Given that a single respiratory cilium generates approximately $60 \ \rm pN$ of force at its tip \citep{hill2010force}, the active force densities generated by our model are of the correct order of magnitude for ciliary carpets. Although the prescribed motions in the translating~\eqref{translating rod} and rotating~\eqref{rotating rod} rod models are symmetric with respect to the Lagrangian label $X$ at the level of individual cilia, the traveling metachronal waves arising from the phase lag between neighboring cilia break translational symmetry in the Eulerian frame. As a result, the force densities are biased, with a larger portion of the forcing being directed in the positive $x$ direction. 

The numerically obtained steady state flow fields at zero applied pressure (figure \ref{fig:flow fields} middle column) reflect this bias, with each model driving a net fluid flux. To quantify this flux, we compute the dimensional mean streamwise velocity profile $u(y)$ and corresponding volumetric flow rate $Q$
\begin{equation}
u(y) = \frac{1}{L}\int_0^L \bm{u}(x,y,t)\cdot \bm{e}_x , dx , \qquad
Q = \int_0^H u(y),dy,
\label{eq:flow rate}
\end{equation}
with $Q$ expressed per unit depth. Streamwise flow profiles for model are plotted in figure \ref{fig:flow fields}(right column). Because the steady-state velocity field is both spatially and temporally periodic, time-averaging over one beat cycle is equivalent to spatial averaging over one metachronal wavelength (appendix~\ref{App:numerics} and figure~\ref{fig:convergence testing}a,b).

Despite sharing the same reference ciliary fraction and metachronal wavelength, as well as identical beat frequency and comparable oscillation amplitudes, the experimentally measured kinematics generate a substantially larger maximal force density, flow velocity, and flow rate compared to the symmetrically oscillating rod models. Importantly, the three cases also exhibit distinct flow topologies. Purely translational oscillations produce little to no recirculation within the channel, whereas rigid rotational oscillations generate recirculation zones in the channel core that are detrimental to net transport. In contrast, the experimentally measured kinematics induce a coherent net flow through the channel core, accompanied by localized recirculation near the wall during the power-stroke phase of the metachronal wave. The latter flow organization is favorable for both efficient transport across the channel and for enhanced mixing within the ciliary layer~\citep{Ding2014}.

These results demonstrate the robustness and flexibility of the model formulation and spectral algorithm proposed here for computing flows in multi-layered ciliated channels across a range of ciliary kinematics. In the following, we focus on the first model of ciliary kinematics, the translating rod model, and analyze it in depth both analytically and numerically.

\begin{figure}
    \centering
    \includegraphics[width=\linewidth]{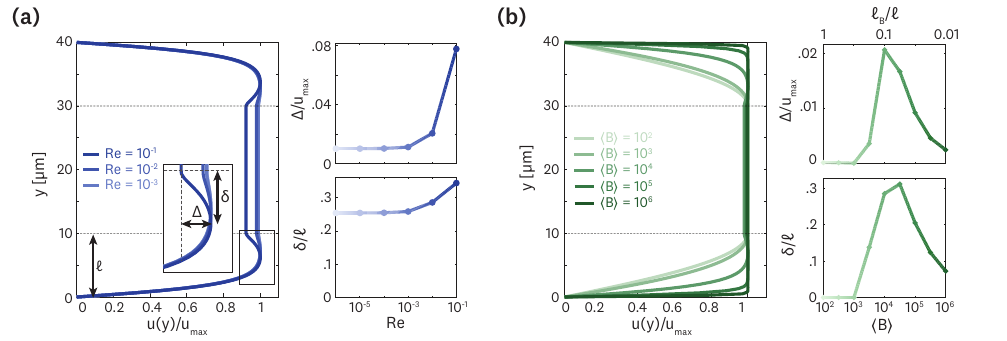}
    \caption{\textbf{Quantifying inertial effects.} Lag magnitude is denoted $\Delta$ and lag penetration depth $\delta$. (a) Period-averaged flow profile, normalized lag magnitude, and normalized penetration depth at fixed $\langle B\rangle = 10^4$ and varied $\rm Re$. Note, $u(y)$ profiles for $\rm{Re} < 10^{-3}$ coincide with the profile for $\rm{Re} = 10^{-3}$. (b) Period-averaged flow profile, normalized lag magnitude, and normalized penetration depth at fixed $\rm Re = 0.01$ and varied $\langle B\rangle$. The lag magnitude and penetration depth are maximized when $\ell_B/\ell \sim 0.1$}
    \label{fig:lag magnitude}
\end{figure}

To conclude this section, we note that the period-averaged flow profiles from the translating-rod model at zero applied pressure exhibit a lag between the ciliary layers and the free layer (figure~\ref{fig:flow fields}a). We quantify this effect by defining the lag magnitude $\Delta$ as the difference between the maximum velocity in the ciliary layer and the velocity at the channel center, and the penetration depth $\delta$ as the distance from the fluid-porous media interface at which the velocity decreases by $1\%$ from its maximum (figure~\ref{fig:lag magnitude}).

To characterize the origin of this lag, we perform simulations in which we vary the Reynolds number $\mathrm{Re}$ and mean Brinkman coefficient $\langle{B}\rangle$ independently. We find that both $\Delta$ and $\delta$ increase monotonically with $\mathrm{Re}$ over the range relevant to ciliary flows, indicating that the lag arises from finite unsteady and inertial effects. In contrast, their dependence on $\langle{B}\rangle$ is non-monotonic: the lag is negligible when the Brinkman screening length $\ell_B$ is comparable to the ciliary layer thickness $\ell$, increases to a maximum at intermediate values satisfying $\ell_B/\ell \sim 0.1$, and decreases again as $\ell_B$ becomes small. This behavior reflects the competition between viscous screening within the porous layer and coupling to the free flow in the inner core: weak screening suppresses the buildup of velocity within the ciliary layer, whereas strong screening dampens interactions with the free layer over short lengthscales. Together, these results show that the observed lag is governed by the interplay between inertial effects and Brinkman screening.

\section{Mean-field model and analytical solution}
\label{sec:linear}

In this section, we derive an analytical approximation of the inverse kinematic mapping $\boldsymbol{\chi}_c^{-1}$ for the translating-rod model~\eqref{translating rod}, and use its period average to obtain a reduced one-dimensional mean-field approximation of~\eqref{NSB}. We then solve the resulting equation analytically, which yields inexpensive predictions of the average flow profile $u(y)$ as a function of the applied pressure, ciliary parameters and channel geometry.

We begin by performing a regular perturbation expansion in the oscillation amplitude $a$ of the translating-rod model \eqref{translating rod}. Retaining terms up to $O(a^2)$ yields
\begin{equation} \label{perturb inv TR}
\bm{\chi}_c^{-1}(x,t) =  X(x,t)
= x
- a \cos(\omega t + kx)
-  a^2 k \cos(\omega t + kx)\sin(\omega t + kx)
+ O(a^3).
\end{equation}
Substituting \eqref{perturb inv TR} into \eqref{current Phi} gives, to second order,
\begin{equation} \label{perturb phi and vc}
\begin{aligned}
B(x,t)
&= \frac{\langle B\rangle}
{1 -  a k \sin(\omega t + kx)
  + a^2 k^2 \cos^2(\omega t + kx)},\\[4pt]
u_c(x,t)
&= -a\omega \sin(\omega t + kx)
   +  a^2 \omega k \cos^2(\omega t + kx),
\end{aligned}
\end{equation}
where $u_c = \bm{v}_c\cdot\bm{e}_x$ is the streamwise component of the ciliary velocity $\bm{v}_c$. 
Averaging over one beat period yields the mean Brinkman coefficient $\bar{B} \coloneq \langle B \rangle$ and mean ciliary speed $\bar{U}_c$, opposite to the direction of metachronal wave propagation,
\begin{equation}
\bar{B} \coloneq \langle B \rangle, 
\qquad
\bar{U}_c\coloneq \langle u_c \rangle 
= \frac{1}{2}    a^2 \omega  k .
\label{eq:perturb B and vc}
\end{equation} 
We find that metachronal waves introduce an $O(a^2)$ drift velocity. In the absence of metachronal coordination ($k=0$), $\bar{U}_c$ vanishes identically.

We now derive a mean-field model for the steady state, period-averaged streamwise flow profile $u(y)$. Exploiting the traveling-wave structure of both the forcing and the steady-state response, we replace the time-dependent Brinkman coefficient $B$ and ciliary velocity $\bm{v}_c$ in~\eqref{NSB} by their temporal (equivalently, spatial) averages $\bar{B} = \langle B \rangle$ and $\bar{U}_c\bm{e}_x$. 
 By virtue of averaging, flow unsteadiness and variations in $x$ are eliminated. Additionally, since by channel symmetry, cross-stream flows do not contribute to the net streamwise transport, we neglect the $y$-component of the flow field. The Navier–Stokes–Brinkman equations in~\eqref{NSB} thus reduce to a steady, one dimensional Stokes–Brinkman boundary-value problem with layered structure,
\begin{equation} \label{1D ODEs}
\begin{cases}
u'' - \bar{B}\big(u-\bar{U}_c\big) - \Delta P = 0,
&\qquad  0 \le y \le \ell ,\\[4pt]
u'' - \Delta P = 0,
&\qquad  \ell \le y \le H-\ell,\\[4pt]
u'' - \bar{B}\big(u-\bar{U}_c\big) - \Delta P = 0.
&\qquad  H-\ell \le y \le H,
\end{cases}
\end{equation}

We impose no-slip at the channel walls and continuity of velocity and shear at the interfaces at $\ell$ and $(H-\ell)$,
\begin{equation}\label{eq:ODEbc}
\begin{aligned}
u(0)& = u(H)=0,\qquad
u({\ell^-})= u({\ell^+}), \qquad
u'(\ell^-)= u'(\ell^+)\\
&  
\left. u\right|_{(H-\ell)^-}=\left. u\right|_{(H-\ell)^+},\quad \quad \left. u'\right|_{(H-\ell)^-}=\left. u'\right|_{(H-\ell)^+}.
\end{aligned}
\end{equation}
Solving~\eqref{1D ODEs} in each layer, enforcing the boundary conditions~\eqref{eq:ODEbc}, and rewriting in terms of the Brinkman screening length $\ell_B$ yields
\begin{equation} 
\label{ode solution}
u(y)=
\begin{cases}
\left(\bar{U}_c- \ell_B^2 \Delta P\right)
\!\left[1-\cosh\dfrac{y}{\ell_B}\right]
+ \beta\,\sinh\dfrac{y}{\ell_B},
& 0 \le y \le \ell,\\[12pt]
\dfrac{1}{2}(\Delta P)\,y^2 - \dfrac{1}{2} (H \Delta P ) y + D, 
& \ell\le y \le H-\ell,\\[12pt]
\left(\bar{U}_c-  \ell_B^2 \Delta P\right)
\!\left[1-\cosh\dfrac{H-y}{\ell_B}\right]
+ \beta\,\sinh\dfrac{H-y}{\ell_B},
& H-\ell \le y \le H,
\end{cases}
\end{equation}
with constants
\begin{equation} \label{ode constants}
\begin{aligned}
\beta &=
\left(\bar{U}_c-\ell_B^2 \Delta P \right)
\tanh\dfrac{\ell}{\ell_B}
-\frac{1}{2} \ell_B (H-2\ell)\Delta P \ {\rm sech}\dfrac{\ell}{\ell_B}, 
\\
D & = \frac{1}{2} \ell(H-\ell)\Delta P+
\left(\bar{U}_c- \ell_B^2 \Delta P\right)
\!\left(1-{\rm sech}\dfrac{\ell}{\ell_B}\right)
-\frac{1}{2} \ell_B
(H-2\ell) \Delta P\ \tanh\dfrac{\ell}{\ell_B}.
\end{aligned}
\end{equation}
The solution $u(y)$ consists of two symmetric, cilia-driven Brinkman layers adjacent to the walls and a central Stokes core. The hyperbolic structure in the outer layers reflects momentum screening over the Brinkman screening length $\ell_B$, while the core exhibits, in addition to the contribution from the ciliary layers, the expected parabolic profile of pressure-driven Stokes flow. We now examine the behavior of this solution in relevant asymptotic limits.


In the absence of the porous layers, for $\bar{B} = 0$, \eqref{1D ODEs} collapses to a reverse Poiseuille flow  over the entire channel width, where flow is only driven by the externally-applied adverse pressure gradient $\Delta P$ (appendix~\ref{App:validate}). Similarly, in the limit where the activity is zero, that is, $\bar{U}_c = 0$, \eqref{ode solution} and \eqref{ode constants} reduce to the solution for flow through a 2D channel with homogeneous, static porous layers lining the top and bottom walls, reminiscent of the analytical solutions derived by~\cite{kuznetsov1996analytical, vafai1990fluid}. Here too, flow is only driven by the externally applied pressure gradient $\Delta P$, and the velocity profile consists of a reverse Poiseuille-like flow profile in the center of the channel, superimposed on the screened flow within the ciliary layers.

Next, we examine the limit of strong Brinkman drag where $\bar{B}$ is large and $\ell_B \ll \ell$, with $\bar{U}_c$ nonzero. Here,~\eqref{ode solution} remains unchanged, but $\tanh(\ell/\ell_B) \to 1$ and sech$(\ell/\ell_B) \to 0$, such that the constants in~\eqref{ode constants} become 
\begin{equation} 
\begin{aligned}
\beta &= \bar{U}_c - \ell_B^2 \Delta P ,
\qquad 
D  =\bar{U}_c + \frac{1}{2} \left[ \ell(H-\ell) +
- 2\ell_B^2 
- \ell_B (H-2\ell) \right]\Delta P
.
\end{aligned}
\end{equation}
In this regime, pressure-driven effects are strongly attenuated within the porous layers due to momentum screening over the short length scale $\ell_B$. As a result, even when the applied pressure gradient is sufficient to reverse the flow in the free core, the flow within the ciliary layers remains largely unaffected; reversing it requires an adverse pressure of order $\ell_B^{-2}$.

In the same strong Brinkman-drag regime, setting $\Delta P = 0$ yields,
\begin{equation} 
u(y)=
\begin{cases}
\bar{U}_c
\!\left[1-\cosh\dfrac{y}{\ell_B}\right]
+ \bar{U}_c\sinh\dfrac{y}{\ell_B},
& 0 \le y \le \ell,\\[8pt]
\bar{U}_c , 
& \ell\le y \le H-\ell,\\[8pt]
\bar{U}_c
\!\left[1-\cosh\dfrac{H-y}{\ell_B}\right]
+ \bar{U}_c\sinh\dfrac{H-y}{\ell_B},
& H-\ell \le y \le H.
\end{cases}
\end{equation}
so that both the peak velocity in the active layers and the velocity in the core attain the ciliary speed $\bar{U}_c$. The resulting profile is plug-like, with a uniform core flow entrained at speed $\bar{U}_c$ and thin boundary layers of thickness $O(\ell_B)$ near the walls.  This behavior is consistent with the flow profile shown in figure~\ref{fig:flow fields}a.


\begin{figure}
    \centering
    \includegraphics[width=\linewidth]{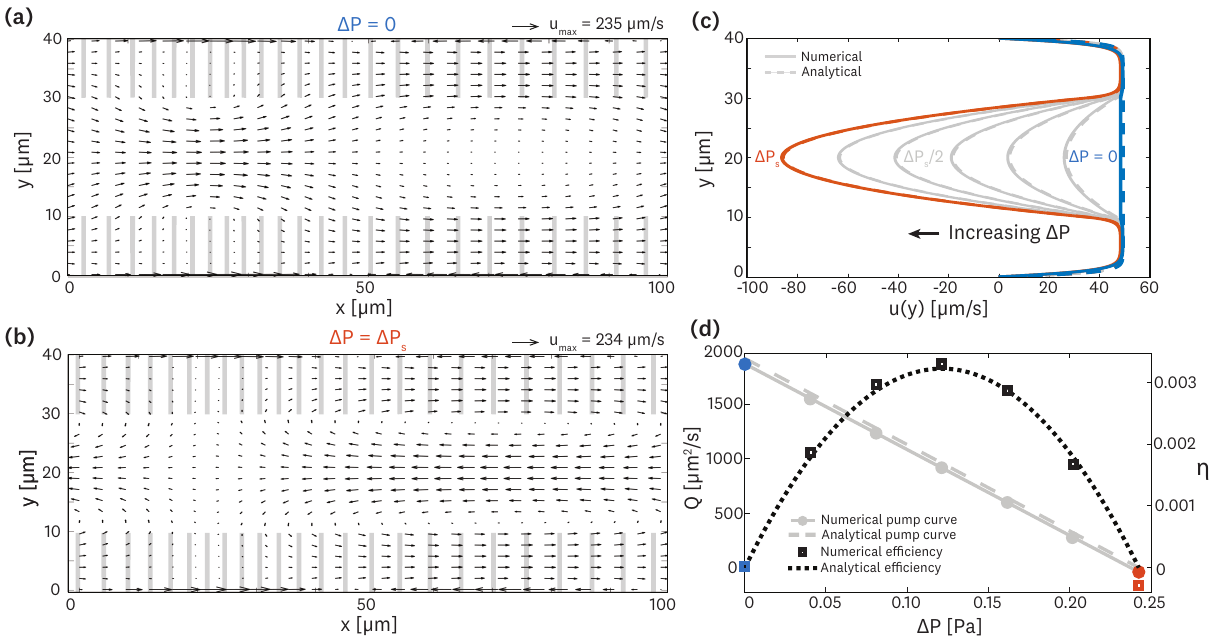} 
    \caption{\textbf{Flow field and pumping performance for fixed channel geometry and ciliary fraction.} (a) Flow field at zero applied pressure $\Delta P = 0$, i.e. the no-load flow. (b) Flow field at stall, $\Delta P = \Delta P_{s}$ with zero net flow $Q = 0$. (c) Period averaged flow profiles with corresponding mean-field analytical predictions (dashed). Blue curves indicate $\Delta P = 0$, orange curves indicate $\Delta P = \Delta P_s$, and gray curves indicate intermediate values of $\Delta P$. (d) Pump curve (linear) and efficiency curve (quadratic). Colored markers indicate no-load (blue) and stall (orange) corresponding with the curves in (c). Here, $\langle \Phi\rangle = 0.2$, $\bar{B} = 3.4\times10^4$, and $H = 40 \ \mu m$, corresponding to $c = 0.5$. For all simulation results plotted here and thereafter, $a = 5 \ \rm \mu m$, $\ \omega = 10 \rm \ Hz$, $d = 0.2 \rm \ \mu m$, $\ell = 10\ \rm \mu m$, $L = 100\ \rm \mu m$, and Re$ = 0.01$.}
    \label{fig:numerical experiment}
\end{figure}

\section{Parametric analysis of ciliary transport}
\label{sec:parametric}

In this section, we examine how the morphology of a ciliated channel governs fluid transport as measured by the flow rate through the channel and the pressure generated by the ciliary layers. Because the flow rate and pressure cannot both be controlled independently, we impose an adverse pressure gradient $\Delta P$ as a way of measuring the pressure generation, while measuring the resulting flow rate $Q$. We begin by fixing the channel morphology and characterizing the pumping response as a function of $\Delta P$, thereby establishing baseline performance metrics. We then vary the mean ciliary fraction $\langle \Phi \rangle$, which controls the total amount of ciliary material in a channel of fixed geometry, and quantify its impact on pumping across the full range of $\Delta P$. Finally, we vary the ciliary confinement ratio $c$, which sets the geometric extent of the ciliary layers, and examine the resulting trade-offs between throughput and pressure generation. This progression isolates the respective roles of ciliary packing and channel geometry in determining transport performance.

To quantify transport performance, we define three complementary metrics. The no-load flow rate $Q_0 = Q|_{\Delta P=0}$ measures the throughput in the absence of an imposed pressure gradient, while the stall pressure $\Delta P_{\rm s} = \Delta P|_{Q=0}$ denotes the maximum adverse pressure that can be sustained before the flow ceases; for $\Delta P > \Delta P_{\rm s}$, the flow reverses and pumping fails. We also define the pumping efficiency $\eta = \mathcal{P}_{\rm out}/\mathcal{P}_{\rm in}$ as the ratio of useful hydraulic power to the power injected by the ciliary layers, where the output power $\mathcal{P}_{\rm out} = Q(\Delta P)\,\Delta P$ is the work performed against the imposed pressure, and the input power $\mathcal{P}_{\rm in} = \int_0^L \!\int_0^H B\, \bm{v}_c \cdot \bm{v}_c \, dx\,dy$ is the rate of mechanical work exerted by the active porous medium. Together, $Q_0$, $\Delta P_{\rm s}$, and $\eta$ quantify the ability of the system to generate flow, sustain pressure, and convert input power into useful transport.

In the mean-field model introduced in section~\ref{sec:linear}, these quantities ($Q_0$, $\Delta P_{\rm s}$, and $\eta$) can be evaluated explicitly. In particular, the input power admits a closed-form expression based on the prescribed kinematics, yielding $\mathcal{P}_{\rm in}= 2\ell L\bar{B}\left({a^2\omega^2}/{2}+{3(a^2\omega k)^2}/{8}\right)$, which provides a direct link between ciliary motion and energetic cost.

We first examine the flow generated by prescribed ciliary activity for a fixed morphology, varying only the imposed pressure $\Delta P$. In figure~\ref{fig:numerical experiment}, representative steady-state velocity fields at no load ($\Delta P=0$) and at stall ($\Delta P=\Delta P_{\rm s}$) are shown in panels (a,b), with the corresponding streamwise period-averaged velocity profiles $u(y)$ in panel (c).

The resulting pump curve $Q$ versus $\Delta P$ is shown in figure~\ref{fig:numerical experiment}d. The pump curve is well described by the linear relation
\begin{equation}
Q(\Delta P) = Q_0\left(1-\frac{\Delta P}{\Delta P_{\rm s}}\right),
\label{pump curve}
\end{equation}
which interpolates between the no-load and stall limits. Next, to evaluate the pumping efficiency $\eta$, we substitute~\eqref{pump curve} into the expression for the output power. We get that $\mathcal{P}_{\rm out} = Q_0 \Delta P \left(1-\Delta P/\Delta P_{\rm s}\right)$ is a quadratic function of $\Delta P$ that vanishes at both no load and stall and attains its maximum at $\Delta P = \Delta P_{\rm s}/2$, corresponding to $Q = Q_0/2$. Because the input power $\mathcal{P}_{\rm in}$ is independent of $\Delta P$, the efficiency $\eta$ is maximized at the same operating point, with $\eta_{\rm max} = Q_0 \Delta P_{\rm s}/(4\mathcal{P}_{\rm in})$. This quadratic dependence of $\eta$ on $\Delta P$, with a maximum at $\Delta P_{\rm s}/2$, is confirmed by the numerical results in figure~\ref{fig:numerical experiment}d.

Results from the linearized analytical solution~\eqref{ode solution}--\eqref{ode constants} are superimposed in figure~\ref{fig:numerical experiment}c,d. For the value of $\bar{B}$ considered here, the mean-field model accurately captures both the velocity profiles and the pump curve, as evidenced by the agreement between analytical (dashed) and numerical (solid) results.

The linear dependence of the flow rate $Q$ on $\Delta P$, together with the corresponding quadratic dependence of the efficiency $\eta$ on $\Delta P$, provides a baseline against which the effects of ciliary packing and confinement are assessed next. 


\begin{figure} [t!]
    \centering
    \includegraphics[width=0.9\linewidth]{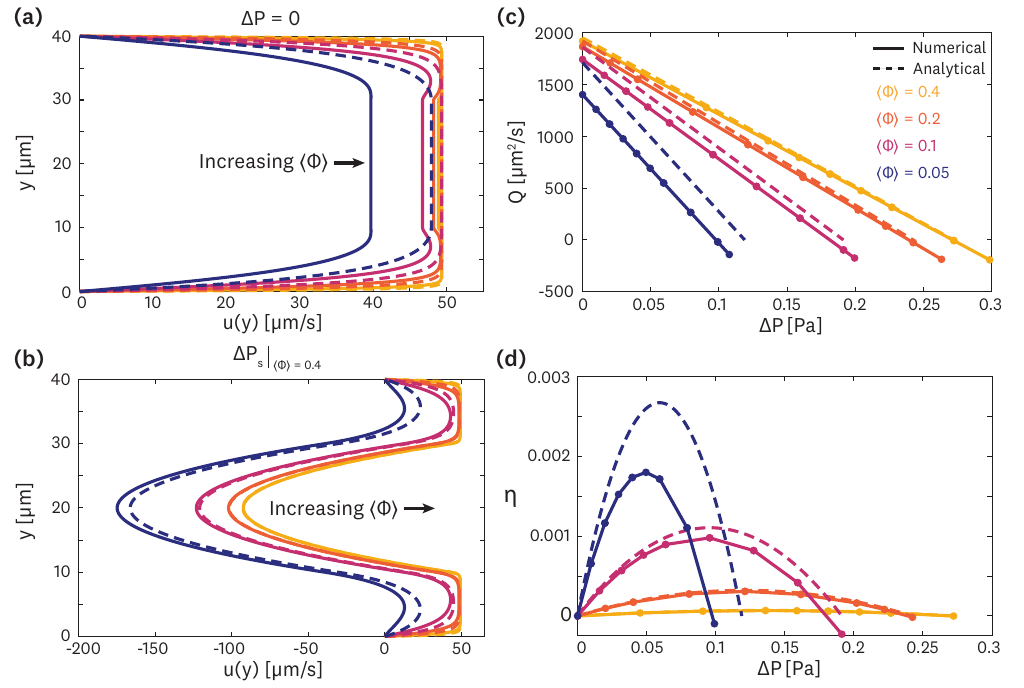}
    \caption{\textbf{Flow profiles and pumping performance for varied ciliary fraction and fixed confinement ratio.} Analytical predictions are indicated with dashed lines, color indicates the ciliary fraction $\langle \Phi\rangle$, with warmer colors indicating more material and higher $\bar{B}$. (a) Period averaged flow profiles at $\Delta P = 0$. (b) Period averaged flow profiles at the stall pressure of the largest $\langle \Phi\rangle$ plotted. (c) Pump curves corresponding with the set of profiles plotted in (a,b). More ciliary material increases both the no-load flow rate and stall pressure. (d) Efficiency curves corresponding with the pump curves in (c). Here $H = 40 \ \rm{\mu m }$ and $c = 0.5$. $\langle \Phi\rangle$ = [0.05, 0.1, 0.2, 0.4] corresponds with nondimensional Brinkman coefficient $\bar{B} = [1.8\times 10^3, 7.6\times 10^3, 3.4 \times 10^4, 1.8\times 10^5]$.}
    \label{fig:vary phi}
\end{figure}

\begin{figure} [!t]
    \centering
    \includegraphics[width=0.9\linewidth]{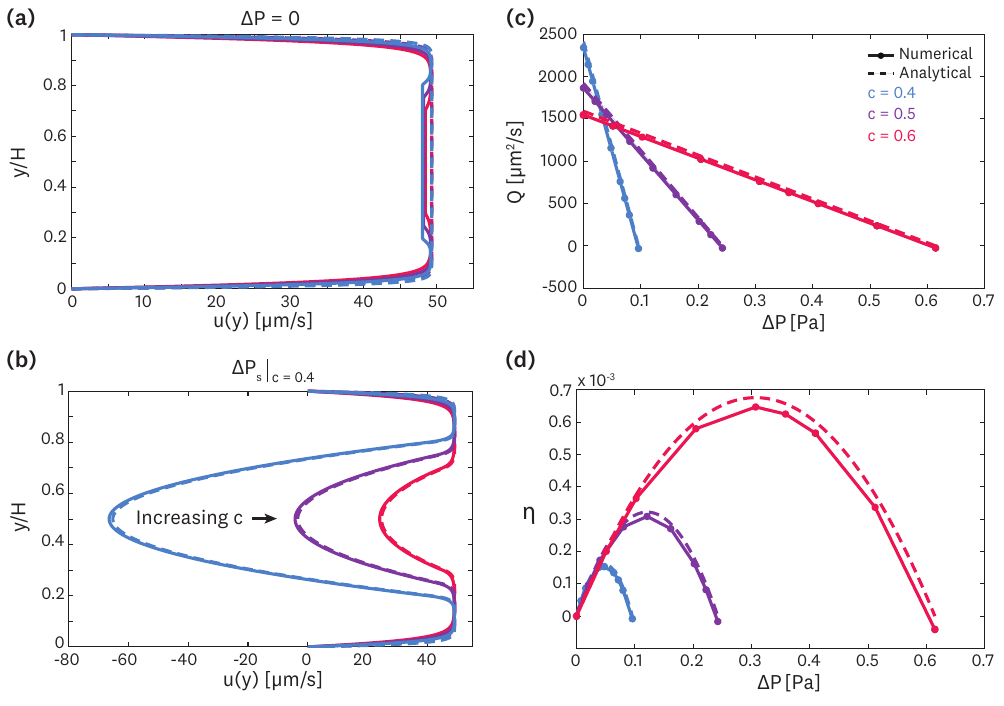}
    \caption{\textbf{Flow profiles and pumping performance for fixed ciliary fraction and varied confinement ratio.} (a) No load flow profiles plotted on normalized y axes. (b) Flow profiles at the stall pressure of the smallest $2 \ell /H$. Increasing the ciliated fraction allows the pumps to sustain larger adverse pressure before flow reversal. (c) Pump curves corresponding with the profiles in (a,b). Larger ciliated fraction, and thus smaller channel height $H$, decreases the no load flow rate and increases the stall pressure. (d) Efficiency curves corresponding with the pump curves in (c). $c = [0.4,0.5,0.6]$ corresponds with $H = [50, 40, 33.33].$ For all panels, ciliary material is fixed at $\langle \Phi\rangle = 0.2$ corresponding with reference Brinkman coefficient $\bar{B} = 3.4\times10^4$.}
    \label{fig:vary H}
\end{figure}

\begin{figure} [t!]
    \centering 
    \includegraphics[width=\linewidth]{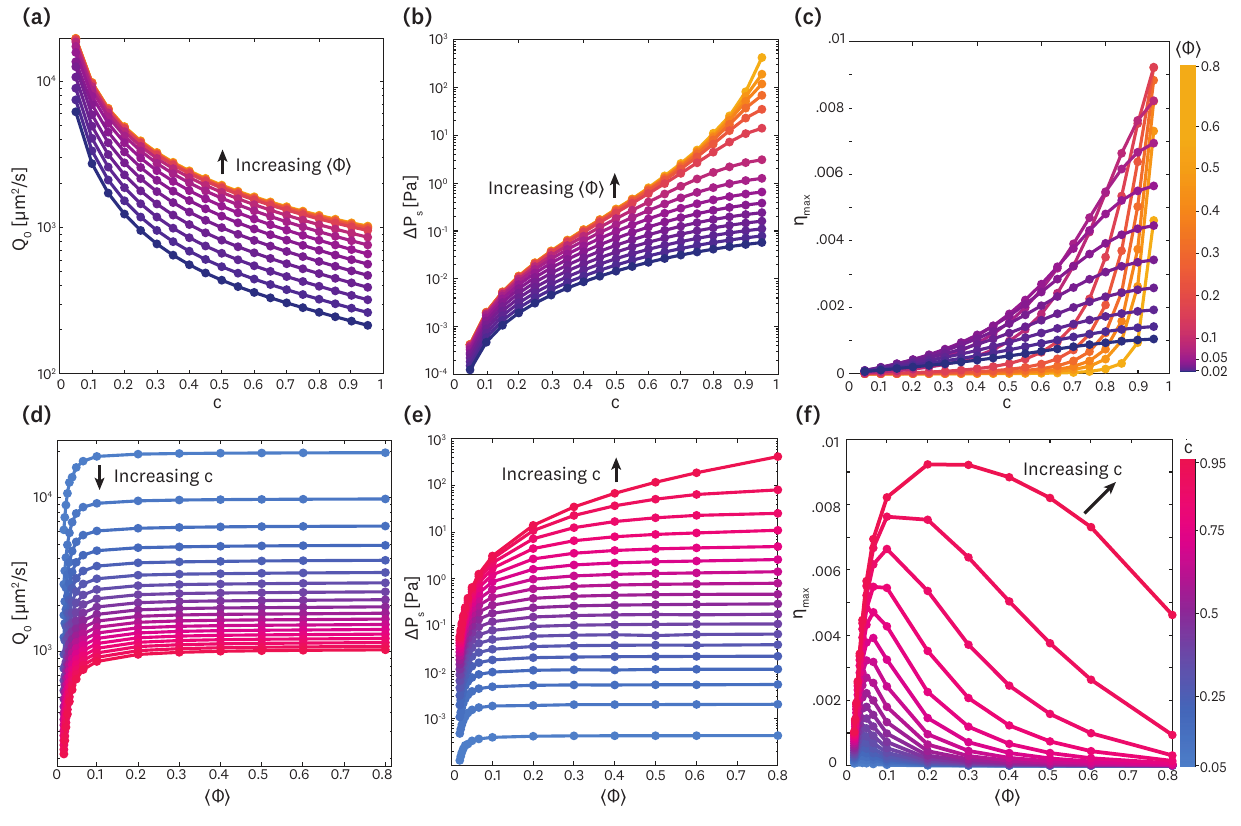}
    \caption{\textbf{Numerical results for varied confinement ratio (a-c) and ciliary fraction (d-f).} (a) No-load flow rate versus confinement ratio. Increased fraction increases the no load flow rate at every confinement ratio. (b) Stall pressure versus confinement ratio. Increased ciliary fraction increases the stall pressure at every confinement ratio. (c) Maximum pumping efficiency versus confinement ratio. The amount of ciliary material non-monotonically influences the maximum pumping efficiency. (d) No-load flow rate versus ciliary fraction $\langle \Phi \rangle$. Increased confinement ratio decreases the no-load flow rate because the channel size shrinks. (e) Stall pressure versus ciliary fraction. Increased confinement ratio increases the stall pressure. (f) Maximum pumping efficiency versus ciliary fraction. For each confinement ratio, there exists and optimal amount of ciliary material for which the maximum pumping efficiency is maximized. $\langle \Phi \rangle = [0.02-0.8]$ corresponds with interciliary spacings $b = [10-0.25] \ \rm{\mu m}$, and $\bar{B} = [1.8 \times 10^2 - 2.2 \times 10^6]$. $c = [0.05-0.95]$ corresponds with $H= [400-21] \ \rm \mu m $.}
    \label{fig:vary phi and H}
\end{figure}

In figure~\ref{fig:vary phi}, we vary the mean ciliary fraction $\langle \Phi \rangle$, which controls the total amount of ciliary material, while keeping the ciliary confinement ratio $c$ fixed. Increasing $\langle \Phi \rangle$ enhances the Brinkman drag within the ciliary layers, thereby reducing the screening length and localizing momentum transfer near the walls. This effect is evident in the period-averaged velocity profiles (figure~\ref{fig:vary phi}a,b), where larger $\langle \Phi \rangle$ produces steeper velocity gradients within the ciliary layers and stronger shear at the interface with the free fluid. As a consequence, both the no-load flow rate and the stall pressure increase with $\langle \Phi \rangle$ (figure~\ref{fig:vary phi}c), reflecting the enhanced ability of the ciliary layer to both drive flow and sustain adverse pressure gradients. However, this increase in transport capacity comes at the expense of efficiency: the pumping efficiency decreases with increasing $\langle \Phi \rangle$ (figure~\ref{fig:vary phi}d), indicating that in this parameter range, denser ciliary packing incurs a higher energetic cost per unit transported fluid. 

Next, we fix the total amount of ciliary material and vary the ciliary confinement ratio $c$ by holding $\ell$ fixed and varying the channel height $H$ (figure~\ref{fig:vary H}). This variation introduces a trade-off between bulk transport and the ability to sustain flow against adverse pressure. At no load (figure~\ref{fig:vary H}a), the flow generated within the ciliary layers entrains fluid in the channel interior even as $H$ increases, leading to an increase in the no-load flow rate as $c$ decreases. In contrast, under an imposed adverse pressure gradient (figure~\ref{fig:vary H}b), channels with smaller $c$ exhibit pronounced backflow in the center region, resulting in a reduced stall pressure. These trends are reflected in the pump curves shown in figure~\ref{fig:vary H}c. Because the total ciliary material is fixed, the input power remains constant across this set of simulations. The increase in stall pressure $\Delta P_{\rm s}$ with increasing $c$ outweighs the corresponding decrease in no-load flow rate $Q_0$, so that the overall pumping efficiency $\eta$ increases with confinement.

In figure~\ref{fig:vary phi and H}, we summarize the numerically computed transport performance over the full range of ciliary confinement ratio $c$ and mean ciliary fraction $\langle \Phi \rangle$. In the top row (figure~\ref{fig:vary phi and H}a–c),
the no-load flow rate $Q_0$, stall pressure $\Delta P_{\rm s}$, and maximum pumping efficiency $\eta_{\rm max}$ are shown as functions of $c$ for different values of $\langle \Phi \rangle$, while the bottom row (figure~\ref{fig:vary phi and H}d-f) presents the same quantities as functions of $\langle \Phi \rangle$ for different values of $c$. 
Each curve in the bottom row therefore corresponds to a vertical slice of the results shown in the top row.

Consistent with the trends identified in figures~\ref{fig:vary phi} and~\ref{fig:vary H}, the no-load flow rate (figure~\ref{fig:vary phi and H}a,d) decreases with increasing confinement $c$ and increases with increasing ciliary fraction $\langle \Phi \rangle$, with the latter dependence saturating for $\langle \Phi \rangle \approx 0.1$–$0.2$. In contrast, the stall pressure increases monotonically with both $c$ and $\langle \Phi \rangle$ (figure~\ref{fig:vary phi and H}b,e), exhibiting a particularly strong sensitivity to confinement: increasing $c$ from $0.05$ to $0.95$ leads to an increase in stall pressure spanning several orders of magnitude. Saturation with respect to $\langle \Phi \rangle$ is observed at moderate confinement but is delayed as $c$ increases.

The maximum pumping efficiency (figure~\ref{fig:vary phi and H}c,f) exhibits a non-monotonic dependence on $\langle \Phi \rangle$. For each $c$ value, there exists an optimal ciliary fraction that maximizes efficiency, resulting in a well-defined peak. As $c$ increases, both the optimal $\langle \Phi \rangle$ and the corresponding maximum efficiency shift to larger values. These results demonstrate that transport performance is governed by a coupled dependence on ciliary fraction and geometric confinement.

\begin{figure} [t]
    \centering
    \includegraphics[width=\linewidth]{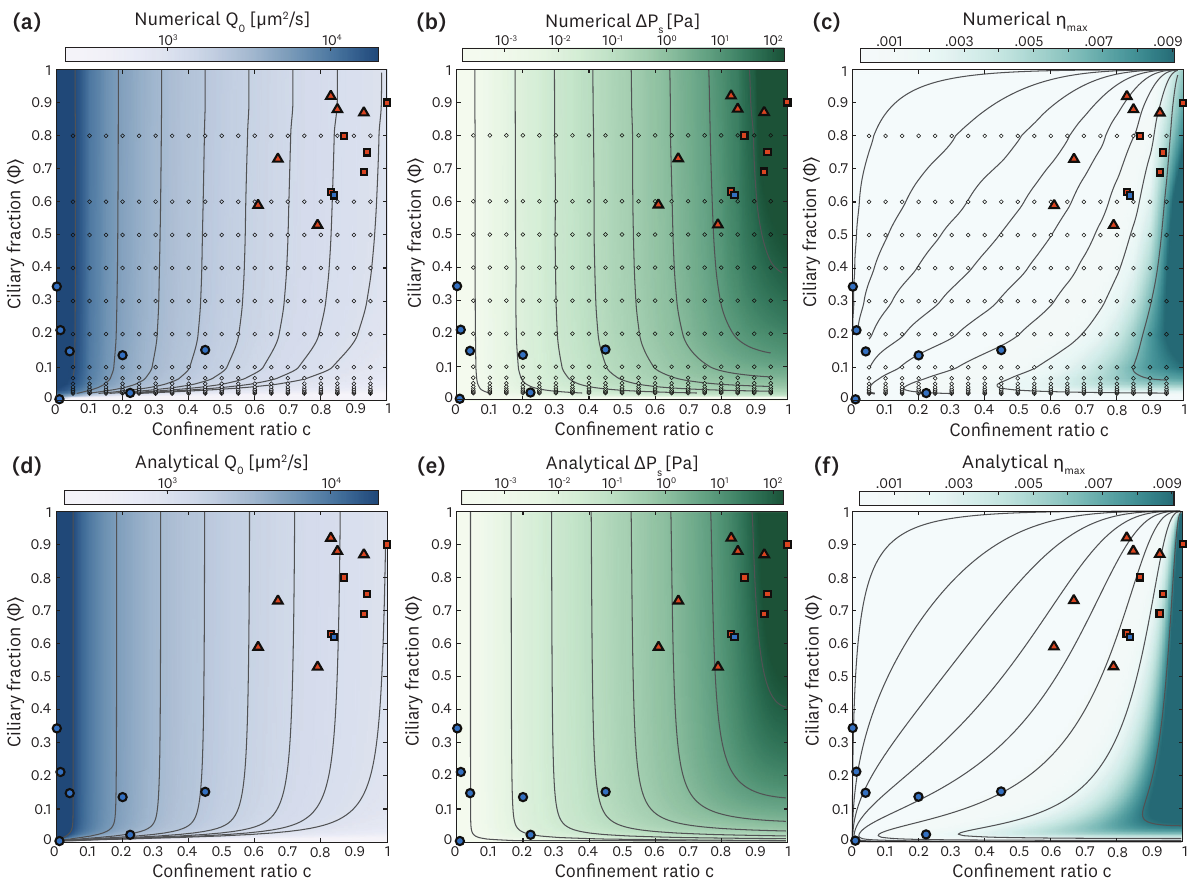}
    \caption{\textbf{Model results overlaid with biological data.} Numerical no-load flow rate (a), stall pressure (b), and maximum pumping efficiency (c). Analytical no-load flow rate (d), stall pressure (e), and maximum pumping efficiency (f). Contour lines in indicate equal elevation and are spaced evenly in percentiles of the data. Colored markers correspond with the biological data in figure \ref{fig:intro} and table \ref{tab:biological data}. Blue markers indicate ciliary carpets and orange markers ciliary flames. Circles represent ducts used for bulk transport, triangles for filtration, and squares for ducts with unknown functions. Small circles in (a-c) indicate numerical simulations, in between the markers the colormaps are interpolated, outside of the simulation region, the colormap is extrapolated.}
    \label{fig:color maps}
\end{figure}

Figure~\ref{fig:color maps} summarizes the numerical and analytical results in the $(c,\langle \Phi \rangle)$ parameter space. The top (analytical) and bottom (numerical) rows report the no-load flow rate, stall pressure, and maximum pumping efficiency. These maps reveal a fundamental trade-off between bulk transport and pressure generation. The no-load flow rate is maximized at small confinement $c$ and intermediate ciliary fraction $\langle \Phi \rangle \approx 0.1$, beyond which further increases in $\langle \Phi \rangle$ yield only marginal gains. In contrast, the stall pressure increases monotonically with both $c$ and $\langle \Phi \rangle$ and is maximized at large confinement and high ciliary fraction. The maximum pumping efficiency is optimized at intermediate $\langle \Phi \rangle \approx 0.1$--$0.2$ and large $c$. These optimal regions do not coincide, indicating that a ciliated channel cannot simultaneously maximize throughput, pressure generation, and efficiency.

The analytical model accurately captures the no-load flow rate, stall pressure, and maximum efficiency across most of parameter space, with deviations exceeding $10\%$ confined to the regime of weak ciliary packing, $\langle \Phi \rangle \lesssim 0.1$, where the assumptions underlying the mean-field approximation break down.

Placing these results in a biological context, the measurements of $(c,\langle \Phi \rangle)$ reported in figure~\ref{fig:intro}c cluster in distinct regions of parameter space. Ciliary carpets lie near the regime that maximizes the no-load flow rate with the minimal amount of ciliary material, while ciliary flames occupy the regime of high stall pressure. This alignment is consistent with their respective physiological roles: carpets are specialized for high-throughput transport in relatively unconstrained environments, whereas flames operate in narrow, resistive ducts where pressure generation is critical for filtration and pumping. Notably, neither class of systems lies near the region of maximal hydrodynamic efficiency, suggesting that energetic optimality is not the primary design constraint in these systems.

Overall, these results demonstrate that ciliary transport is governed by a fundamental geometric and material trade-off: low confinement and moderate ciliary fraction favor throughput, whereas strong confinement and dense ciliary packing favor pressure generation. This trade-off provides a unifying physical framework for interpreting the diversity of ciliated transport systems observed in biology.


These results also raise a natural question: why not combine the advantages of both transport regimes by constructing a large-diameter ciliary flame with high ciliary confinement and high ciliary fraction? Such a configuration could, in principle, exploit the enhanced throughput associated with large ducts while simultaneously maintaining the strong pressure generation characteristic of densely packed ciliary layers. A ciliary flame, however, requires sufficiently long and densely packed cilia to span and effectively fill the duct cross-section (figure \ref{fig:intro}b). Because cilia length is biologically constrained, flame-like morphologies can at best be sustained in ducts whose diameter is comparable to the cilia length, namely $H \leq \ell$. This explains why ciliary flames in nature tend to occur in narrow ducts lined with long, densely packed cilia and suggests that the trade-off between throughput and pressure generation identified here arises, at least in part, from biological constraints on achievable cilia length and packing density -- constraints that could potentially be relaxed in engineered active transport systems.
\section{Discussion}
\label{sec:discuss}


In this work, we developed a unified theoretical framework for cilia-driven transport in confined channels by modeling ciliary layers as an active porous medium governed by a Navier–Stokes–Brinkman formulation. This approach bridges the gap between classical envelope models, which treat the ciliary layer as an impermeable boundary~\citep{Blake1971,Blake1972}, and filament-resolved simulations that explicitly capture individual cilia dynamics and hydrodynamic interactions~\citep{Ding2014,Guo2014,Nawroth2017,Man2020, man2020multisynchrony}. By introducing a coarse-graining procedure that, for prescribed ciliary kinematics, links the mean ciliary fraction to an effective Brinkman drag, we identified a minimal set of parameters—namely, the ciliary confinement ratio $c$ and the mean ciliary fraction $\langle \Phi \rangle$—that govern transport. Within this framework, we showed that ciliary pumping is characterized by a linear pressure–flow relation, a quadratic efficiency curve, and a fundamental trade-off between throughput and pressure generation, consistent with mean-field descriptions of active flows~\citep{ling2024flow}.

Exploring this parameter space—the space of ciliary confinement ratio $c$ and mean ciliary fraction $\langle \Phi \rangle$—revealed that low confinement and moderate ciliary fraction favor high-throughput transport, whereas strong confinement and dense ciliary packing enhance the ability to sustain adverse pressure gradients, consistent with predictions from a simpler model in~\cite{ling2024flow}. These regimes align with biological observations: ciliary carpets, which operate in relatively wide lumens and exhibit coordinated metachronal waves, cluster in the high-throughput regime~\citep{kanale2022spontaneous,ling2024flow}, while ciliary flames, which function in narrow, resistive ducts, lie in the pressure-generating regime~\citep{ling2024flow}. This correspondence supports the view that the diversity of ciliary architectures reflects underlying physical constraints imposed by flow and geometry, rather than a single universal optimal design.

While our results align with available biological data, several aspects of ciliated transport remain to be incorporated. The present model reduces geometric and material complexity to two effective parameters, enabling a clear connection between morphology and function, but does not capture the full structural richness of biological systems~\citep{Ramirez2020,ling2024flow}. Many ciliated ducts exhibit spatially varying diameters, curvature, and patchy ciliation, with coverage fractions that vary along the organ length~\citep{Ramirez2020}. In addition, some systems transport suspended particles or interact with complex, non-Newtonian fluids such as mucus, and all involve mechanical feedback between the cilia and the surrounding fluid~\citep{Ding2014,Nawroth2017,roth2025structure}. These effects are neglected here but are expected to play an important role in shaping transport in specific physiological contexts. A key advantage of the active porous media framework is that such features can be incorporated systematically, allowing increasing biological realism while maintaining computational tractability.

Beyond ciliated systems, the physical mechanisms identified here may extend more broadly to biological transport processes driven by coordinated active motion. In particular, recent work on plant–fungal symbiosis has shown that traveling-wave-like strategies can regulate nutrient exchange across interconnected networks~\citep{oyarte2025travelling}. Despite the different biological context, these systems share key features with ciliary transport, including spatially distributed forcing, wave-like coordination, and flows in geometric confinement. Similar principles have also been identified in intracellular flows driven by active filament assemblies, where geometry and activity jointly regulate large-scale transport~\citep{Stein2019,dutta2024self,chakrabarti2024cytoplasmic, jain2025geometric}. This suggests that the trade-off between throughput and resistance identified here may represent a general organizing principle governing transport in active biological media.

These findings also have implications for the design of bio-inspired microfluidic pumps~\citep{gauger2009fluid, vilfan2010self, den2013microfluidic,ul2022microscopic, hanasoge2018, liu20263d}. In particular, the identified trade-off between throughput and pressure generation provides a clear guideline for tuning device performance: low confinement and moderate actuation density favor high flow rates, whereas strong confinement and dense actuation enhance pressure generation. This suggests that different operating regimes can be targeted by adjusting geometric confinement and the effective actuation of the driving elements, without requiring detailed control at the level of individual actuators.

Taken together, our results provide a minimal physical framework for understanding how geometry and active forcing jointly determine transport in confined systems~\citep{ling2024flow}. By identifying a small set of governing parameters and the trade-offs they impose, this work offers a unifying perspective on cilia-driven transport and highlights broader connections to other forms of biologically mediated flow, from epithelial ciliary carpets to intracellular streaming and symbiotic transport networks~\citep{kanale2022spontaneous,dutta2024self,oyarte2025travelling}.

\section*{Acknowledgments}
The authors thank Michael J Shelley for proposing to model ciliary beds as oscillating porous media, and David Stein, Janna Nawroth, and Suraj Shankar for insightful discussions. 
Funding support provided by the NSF grants RAISE IOS-2034043 and CBET-2100209, ONR grants N00014-22-1-2655 and N00014-19-1-2035, and NIH grant R01-HL153622 (all to E.K.).  The work of E.K. was supported in part by grant NSF PHY-2309135 to the Kavli Institute for Theoretical Physics (KITP) and by Princeton University through the William R. Kenan, Jr., Visiting Professorship.


\bibliographystyle{jfm}
\bibliography{jfm}

@article{hill2010force,
  title={Force generation and dynamics of individual cilia under external loading},
  author={Hill, David B and Swaminathan, Vinay and Estes, Ashley and Cribb, Jeremy and O'Brien, E Timothy and Davis, C William and Superfine, R},
  journal={Biophysical journal},
  volume={98},
  number={1},
  pages={57--66},
  year={2010},
  publisher={Elsevier}
}

@article{liu20263d,
  title={3D-printed low-voltage-driven ciliary hydrogel microactuators},
  author={Liu, Zemin and Wang, Che and Ren, Ziyu and Wang, Chunxiang and Wang, Wenkang and Ko, Jongkuk and Song, Shanyuan and Hong, Chong and Chen, Xi and Wang, Hongguang and others},
  journal={Nature},
  pages={1--9},
  year={2026},
  publisher={Nature Publishing Group UK London}
}

@article{varga2018lymphatic,
  title={Lymphatic lacunae of the mucosal folds of human uterine tubes—A rediscovery of forgotten structures and their possible role in reproduction},
  author={Varga, Ivan and Kachl{\'\i}k, David and {\v{Z}}i{\v{s}}kov{\'a}, Marianna and Miko, Michal},
  journal={Annals of Anatomy-Anatomischer Anzeiger},
  volume={219},
  pages={121--128},
  year={2018},
  publisher={Elsevier}
}

@article{mobjerg2004morphology,
  title={Morphology of the kidney in the West African caecilian, Geotrypetes seraphini (Amphibia, Gymnophiona, Caeciliidae)},
  author={M{\o}bjerg, N and Jespersen, {\AA} and Wilkinson, M},
  journal={Journal of morphology},
  volume={262},
  number={2},
  pages={583--607},
  year={2004},
  publisher={Wiley Online Library}
}

@article{bunke1994ultrastructure,
  title={Ultrastructure of the metanephridial system in Aeolosoma bengalense (Annelida)},
  author={Bunke, D},
  journal={Zoomorphology},
  volume={114},
  number={4},
  pages={247--258},
  year={1994},
  publisher={Springer}
}

@article{furuya1997fine,
  title={Fine structure of dicyemid mesozoans, with special reference to cell junctions},
  author={Furuya, Hidetaka and Tsuneki, Kazuhiko and Koshida, Yutaka},
  journal={Journal of Morphology},
  volume={231},
  number={3},
  pages={297--305},
  year={1997},
  publisher={Wiley Online Library}
}

@article{purschke1996dorsolateral,
  title={Dorsolateral ciliary folds in the polychaete foregut: structure, prevalence and phylogenetic significance},
  author={Purschke, G{\"u}nter and Tzetlin, Alexander B},
  journal={Acta Zoologica},
  volume={77},
  number={1},
  pages={33--49},
  year={1996},
  publisher={Wiley Online Library}
}

@article{raidt2015ciliary,
  title={Ciliary function and motor protein composition of human fallopian tubes},
  author={Raidt, Johanna and Werner, Claudius and Menchen, Tabea and Dougherty, Gerard W and Olbrich, Heike and Loges, Niki T and Schmitz, Ralf and Pennekamp, Petra and Omran, Heymut},
  journal={Human Reproduction},
  volume={30},
  number={12},
  pages={2871--2880},
  year={2015},
  publisher={Oxford University Press}
}

@article{nanjundappa2019regulation,
  title={Regulation of cilia abundance in multiciliated cells},
  author={Nanjundappa, Rashmi and Kong, Dong and Shim, Kyuhwan and Stearns, Tim and Brody, Steven L and Loncarek, Jadranka and Mahjoub, Moe R},
  journal={Elife},
  volume={8},
  pages={e44039},
  year={2019},
  publisher={eLife Sciences Publications, Ltd}
}

@article{olstad2019ciliary,
  title={Ciliary beating compartmentalizes cerebrospinal fluid flow in the brain and regulates ventricular development},
  author={Olstad, Emilie W and Ringers, Christa and Hansen, Jan N and Wens, Adinda and Brandt, Cecilia and Wachten, Dagmar and Yaksi, Emre and Jurisch-Yaksi, Nathalie},
  journal={Current Biology},
  volume={29},
  number={2},
  pages={229--241},
  year={2019},
  publisher={Elsevier}
}

@article{kazanis2012number,
  title={The number of stem cells in the subependymal zone of the adult rodent brain is correlated with the number of ependymal cells and not with the volume of the niche},
  author={Kazanis, Ilias and Ffrench-Constant, Charles},
  journal={Stem Cells and Development},
  volume={21},
  number={7},
  pages={1090--1096},
  year={2012},
  publisher={SAGE Publications Sage CA: Los Angeles, CA}
}

@article{haemmerle2015neural,
  title={The neural elements in the lining of the ventricular-subventricular zone: making an old story new by high-resolution scanning electron microscopy},
  author={Haemmerle, Carlos Alexandre dos Santos and Nogueira, Maria In{\^e}s and Watanabe, Ii-sei},
  journal={Frontiers in neuroanatomy},
  volume={9},
  pages={134},
  year={2015},
  publisher={Frontiers Media SA}
}

@article{boucher2024high,
  title={High throughput screening of airway constriction in mouse lung slices},
  author={Boucher, Magali and Henry, Cyndi and G{\'e}linas, Louis and Packwood, Rosalie and Rojas-Ruiz, Andr{\'e}s and Fereydoonzad, Liah and Graham, Percival and Boss{\'e}, Ynuk},
  journal={Scientific Reports},
  volume={14},
  number={1},
  pages={20133},
  year={2024},
  publisher={Nature Publishing Group UK London}
}

@article{li2021identification,
  title={Identification of cilia in different mouse tissues},
  author={Li, Xinhua and Yang, Shuting and Deepk, Vishwa and Chinipardaz, Zahra and Yang, Shuying},
  journal={Cells},
  volume={10},
  number={7},
  pages={1623},
  year={2021},
  publisher={MDPI}
}

@article{storch1989internal,
  title={Internal anatomy of Meiopriapulus fijiensis (Priapulida)},
  author={Storch, Volker and Higgins, Robert P and Morse, M Patricia},
  journal={Transactions of the American Microscopical Society},
  pages={245--261},
  year={1989},
  publisher={JSTOR}
}

@article{kato2011ultrastructure,
  title={Ultrastructure and phylogenetic significance of the head kidneys in Thalassema thalassemum (Thalassematinae, Echiura)},
  author={Kato, Chiharu and Lehrke, Janina and Quast, Bj{\"o}rn},
  journal={Zoomorphology},
  volume={130},
  number={2},
  pages={97--106},
  year={2011},
  publisher={Springer}
}

@article{rohde1992ultrastructure,
  title={Ultrastructure of the flame bulbs and protonephridial capillaries of Artioposthia sp.(Platyhelminthes, Tricladida, Geoplanidae)},
  author={Rohde, Klaus and Watson, NA},
  journal={Acta zoologica},
  volume={73},
  number={4},
  pages={231--236},
  year={1992},
  publisher={Wiley Online Library}
}

@article{schramm1978excretory,
  title={On the excretory system of the rotifer Habrotrocha rosa Donner},
  author={Schramm, Uda},
  journal={Cell and Tissue Research},
  volume={189},
  number={3},
  pages={515--523},
  year={1978},
  publisher={Springer}
}

@article{von1962erorterung,
  author    = {Georg K{\"u}mmel},
  title     = {{Und Er{\"o}rterung des Begriffes ``Zelltyp''}},
  journal   = {Cell and Tissue Research},
  volume    = {57},
  pages     = {172--201},
  year      = {1962},
  publisher = {Springer-Verlag}
}

@article{holmberg1982ciliated,
  title={The ciliated brain duct of Oikopleura dioica (Tunicata, Appendicularia)},
  author={Holmberg, Kaj},
  journal={Acta Zoologica},
  volume={63},
  number={2},
  pages={101--109},
  year={1982},
  publisher={Wiley Online Library}
}

@article{kuznetsov1996analytical,
  title={Analytical investigation of the fluid flow in the interface region between a porous medium and a clear fluid in channels partially filled with a porous medium},
  author={Kuznetsov, AV},
  journal={Applied scientific research},
  volume={56},
  number={1},
  pages={53--67},
  year={1996},
  publisher={Springer}
}

@article{vafai1990fluid,
  title={Fluid mechanics of the interface region between a porous medium and a fluid layer—an exact solution},
  author={Vafai, K and Kim, SungJin},
  journal={International Journal of Heat and Fluid Flow},
  volume={11},
  number={3},
  pages={254--256},
  year={1990},
  publisher={Elsevier}
}

@article{osterman2011finding,
  title={Finding the ciliary beating pattern with optimal efficiency},
  author={Osterman, Natan and Vilfan, Andrej},
  journal={Proceedings of the National Academy of Sciences},
  volume={108},
  number={38},
  pages={15727--15732},
  year={2011},
  publisher={National Academy of Sciences}
}

@article{man2020multisynchrony,
  title={Multisynchrony in active microfilaments},
  author={Man, Yi and Kanso, Eva},
  journal={Physical Review Letters},
  volume={125},
  number={14},
  pages={148101},
  year={2020},
  publisher={APS}
}

@article{liu2025flow,
  title={Flow physics of nutrient transport drives functional design of ciliates},
  author={Liu, Jingyi and Costello, John H and Kanso, Eva},
  journal={Nature Communications},
  volume={16},
  number={1},
  pages={4154},
  year={2025},
  publisher={Nature Publishing Group UK London}
}

@article{liu2025optimal,
  title={Optimal feeding in swimming and attached ciliates},
  author={Liu, Jingyi and Man, Yi and Costello, John H and Kanso, Eva},
  journal={Journal of Fluid Mechanics},
  volume={1003},
  pages={A26},
  year={2025},
  publisher={Cambridge University Press}
}

@article{liu2026feeding,
  title={Feeding rates in sessile versus motile ciliates are hydrodynamically equivalent},
  author={Liu, Jingyi and Man, Yi and Costello, John H and Kanso, Eva},
  journal={eLife},
  volume={13},
  pages={RP99003},
  year={2026},
  publisher={eLife Sciences Publications, Ltd}
}

@article{liu2025nutrient,
  title={Nutrient transport in concentration gradients},
  author={Liu, Jingyi and Man, Yi and Kanso, Eva},
  journal={Physical Review Fluids},
  volume={10},
  number={9},
  pages={093104},
  year={2025},
  publisher={APS}
}

@article{roth2025structure,
  title={Structure and function relationships of mucociliary clearance in human and rat airways},
  author={Roth, Doris and {\c{S}}ahin, Ay{\c{s}}e Tu{\u{g}}{\c{c}}e and Ling, Feng and Tepho, Niels and Senger, Christiana N and Quiroz, Erik J and Calvert, Ben A and van der Does, Anne M and G{\"u}ney, Tankut G and Glasl, Sarah and others},
  journal={Nature communications},
  volume={16},
  number={1},
  pages={2446},
  year={2025},
  publisher={Nature Publishing Group UK London}
}

@article{wuttanachamsri2020effects,
  title={Effects of cilia movement on fluid velocity: II numerical solutions over a fixed domain},
  author={Wuttanachamsri, Kanognudge and Schreyer, Lynn},
  journal={Transport in Porous Media},
  volume={134},
  number={2},
  pages={471--489},
  year={2020},
  publisher={Springer}
}

@article{wuttanachamsri2021effects,
  title={Effects of cilia movement on fluid velocity: I model of fluid flow due to a moving solid in a porous media framework},
  author={Wuttanachamsri, Kanognudge and Schreyer, Lynn},
  journal={Transport in Porous Media},
  volume={136},
  number={2},
  pages={699--714},
  year={2021},
  publisher={Springer}
}

@article{sanderson1981ciliary,
  title={Ciliary activity of cultured rabbit tracheal epithelium: beat pattern and metachrony},
  author={Sanderson, Michael J and Sleigh, Michael A},
  journal={Journal of cell science},
  volume={47},
  number={1},
  pages={331--347},
  year={1981},
  publisher={The Company of Biologists Ltd}
}

@article{gilpin2020multiscale,
  title={The multiscale physics of cilia and flagella},
  author={Gilpin, William and Bull, Matthew Storm and Prakash, Manu},
  journal={Nature Reviews Physics},
  volume={2},
  number={2},
  pages={74--88},
  year={2020},
  publisher={Nature Publishing Group UK London}
}

@article{dutta2024self,
  title={Self-organized intracellular twisters},
  author={Dutta, Sayantan and Farhadifar, Reza and Lu, Wen and Kabacao{\u{g}}lu, Gokberk and Blackwell, Robert and Stein, David B and Lakonishok, Margot and Gelfand, Vladimir I and Shvartsman, Stanislav Y and Shelley, Michael J},
  journal={Nature Physics},
  volume={20},
  number={4},
  pages={666--674},
  year={2024},
  publisher={Nature Publishing Group UK London}
}

@article{wei2021measurements,
  title={Measurements of the unsteady flow field around beating cilia},
  author={Wei, Da and Dehnavi, Parviz G and Aubin-Tam, Marie-Eve and Tam, Daniel},
  journal={Journal of Fluid Mechanics},
  volume={915},
  pages={A70},
  year={2021},
  publisher={Cambridge University Press}
}

@article{fulford1986muco,
  title={Muco-ciliary transport in the lung},
  author={Fulford, Glenn R and Blake, John R},
  journal={Journal of theoretical Biology},
  volume={121},
  number={4},
  pages={381--402},
  year={1986},
  publisher={Elsevier}
}

@book{kutz2013data,
  title={Data-driven modeling \& scientific computation: methods for complex systems \& big data},
  author={Kutz, Jose Nathan},
  year={2013},
  publisher={OUP Oxford}
}

@article{kanale2022spontaneous,
  title={Spontaneous phase coordination and fluid pumping in model ciliary carpets},
  author={Kanale, Anup V and Ling, Feng and Guo, Hanliang and F{\"u}rthauer, Sebastian and Kanso, Eva},
  journal={Proceedings of the National Academy of Sciences},
  volume={119},
  number={45},
  pages={e2214413119},
  year={2022},
  publisher={National Academy of Sciences}
}

@article{ling2024flow,
  title={Flow physics guides morphology of ciliated organs},
  author={Ling, Feng and Essock-Burns, Tara and McFall-Ngai, Margaret and Katija, Kakani and Nawroth, Janna C and Kanso, Eva},
  journal={Nature physics},
  volume={20},
  number={10},
  pages={1679--1686},
  year={2024},
  publisher={Nature Publishing Group UK London}
}

@article{hanasoge2018,
  title={Microfluidic pumping using artificial magnetic cilia},
  author={Hanasoge, Srinivas and Hesketh, Peter J and Alexeev, Alexander},
  journal={Microsystems \& nanoengineering},
  volume={4},
  number={1},
  pages={1--9},
  year={2018},
  publisher={Nature Publishing Group}
}

@article{Mckanna1968,
  title={Fine structure of the protonephridial system in Planaria},
  author={McKanna, James A},
  journal={Zeitschrift f{\"u}r Zellforschung und Mikroskopische Anatomie},
  volume={92},
  number={4},
  pages={509--523},
  year={1968},
  publisher={Springer}
}

@article{Wei2019,
  title={Is the zero Reynolds number approximation valid for ciliary flows?},
  author={Wei, Da and Dehnavi, Parviz Ghoddoosi and Aubin-Tam, Marie-Eve and Tam, Daniel},
  journal={Physical review letters},
  volume={122},
  number={12},
  pages={124502},
  year={2019},
  publisher={APS}
}

@article{Blake1982,
  title={Flow patterns around ciliated microorganisms and in ciliated ducts},
  author={Blake, JR and Liron, N and Aldis, GK},
  journal={Journal of theoretical biology},
  volume={98},
  number={1},
  pages={127--141},
  year={1982},
  publisher={Elsevier}
}

@article{Gueron1992,
  title={Ciliary motion modeling, and dynamic multicilia interactions},
  author={Gueron, Shay and Liron, Nadav},
  journal={Biophysical journal},
  volume={63},
  number={4},
  pages={1045--1058},
  year={1992},
  publisher={Elsevier}
}

@article{Liron1976,
  title={The discrete-cilia approach to propulsion of ciliated micro-organisms},
  author={Liron, N and Mochon, S},
  journal={Journal of Fluid Mechanics},
  volume={75},
  number={3},
  pages={593--607},
  year={1976},
  publisher={Cambridge University Press}
}

@article{Liron1978,
  title={Fluid transport by cilia between parallel plates},
  author={Liron, N},
  journal={Journal of Fluid Mechanics},
  volume={86},
  number={4},
  pages={705--726},
  year={1978},
  publisher={Cambridge University Press}
}

@article{Stein2019,
  title={Coarse graining the dynamics of immersed and driven fiber assemblies},
  author={Stein, David B and Shelley, Michael J},
  journal={Physical Review Fluids},
  volume={4},
  number={7},
  pages={073302},
  year={2019},
  publisher={APS}
}

@article{Guo2014,
  title={Cilia beating patterns are not hydrodynamically optimal},
  author={Guo, Hanliang and Nawroth, Janna and Ding, Yang and Kanso, Eva},
  journal={Physics of Fluids},
  volume={26},
  number={9},
  pages={091901},
  year={2014},
  publisher={AIP}
}

@article{Faubel2016,
  title={Cilia-based flow network in the brain ventricles},
  author={Faubel, Regina and Westendorf, Christian and Bodenschatz, Eberhard and Eichele, Gregor},
  journal={Science},
  volume={353},
  number={6295},
  pages={176--178},
  year={2016},
  publisher={American Association for the Advancement of Science}
}

@article{Greenstone1985,
  title={Ciliary function in health and disease},
  author={Greenstone, M and Cole, PJ},
  journal={British journal of diseases of the chest},
  volume={79},
  pages={9--26},
  year={1985},
  publisher={Elsevier}
}

@article{Afzelius1976,
  title={A human syndrome caused by immotile cilia},
  author={Afzelius, Bj{\"o}rn A},
  journal={Science},
  volume={193},
  number={4250},
  pages={317--319},
  year={1976},
  publisher={American Association for the Advancement of Science}
}

@article{Michelin2011,
  title={Optimal feeding is optimal swimming for all P{\'e}clet numbers},
  author={Michelin, S{\'e}bastien and Lauga, Eric},
  journal={Physics of Fluids},
  volume={23},
  number={10},
  pages={101901},
  year={2011},
  publisher={American Institute of Physics}
}

@article{Smith2008,
  title={Modelling mucociliary clearance},
  author={Smith, David J and Gaffney, Eamonn A and Blake, John R},
  journal={Respiratory physiology \& neurobiology},
  volume={163},
  number={1},
  pages={178--188},
  year={2008},
  publisher={Elsevier}
}

@article{Ding2014,
  title={Mixing and transport by ciliary carpets: a numerical study},
  author={Ding, Yang and Nawroth, Janna C and McFall-Ngai, Margaret J and Kanso, Eva},
  journal={Journal of Fluid Mechanics},
  volume={743},
  pages={124--140},
  year={2014},
  publisher={Cambridge University Press}
}

@article{Guo2020,
  title={Simulating cilia-driven mixing and transport in complex geometries},
  author={Guo, Hanliang and Zhu, Hai and Veerapaneni, Shravan},
  journal={Physical Review Fluids},
  volume={5},
  number={5},
  pages={053103},
  year={2020},
  publisher={APS}
}

@article{Man2020,
  title={Cilia oscillations},
  author={Man, Yi and Ling, Feng and Kanso, Eva},
  journal={Philosophical Transactions of the Royal Society B},
  volume={375},
  number={1792},
  pages={20190157},
  year={2020},
  publisher={The Royal Society}
}

@article{Nawroth2017,
  title={Motile cilia create fluid-mechanical microhabitats for the active recruitment of the host microbiome},
  author={Nawroth, Janna C and Guo, Hanliang and Koch, Eric and Heath-Heckman, Elizabeth AC and Hermanson, John C and Ruby, Edward G and Dabiri, John O and Kanso, Eva and McFall-Ngai, Margaret},
  journal={Proceedings of the National Academy of Sciences},
  volume={114},
  number={36},
  pages={9510--9516},
  year={2017},
  publisher={National Acad Sciences}
}

@article{Ramirez2020,
  title={Multi-scale spatial heterogeneity enhances particle clearance in airway ciliary arrays},
  author={Ramirez-San Juan, Guillermina R and Mathijssen, Arnold JTM and He, Mu and Jan, Lily and Marshall, Wallace and Prakash, Manu},
  journal={Nature Physics},
  pages={1--7},
  year={2020},
  publisher={Nature Publishing Group}
}

@article{Baeumler2012,
  title={Development of the excretory system in a polyplacophoran mollusc: stages in metanephridial system development},
  author={Baeumler, Natalie and Haszprunar, Gerhard and Ruthensteiner, Bernhard},
  journal={Frontiers in zoology},
  volume={9},
  number={1},
  pages={23},
  year={2012},
  publisher={Springer}
}

@article{Ott2016,
  title={Pronephric tubule morphogenesis in zebrafish depends on Mnx mediated repression of irx1b within the intermediate mesoderm},
  author={Ott, Elisabeth and Wendik, Bj{\"o}rn and Srivastava, Monika and Pacho, Frederic and T{\"o}chterle, Sonja and Salvenmoser, Willi and Meyer, Dirk},
  journal={Developmental Biology},
  volume={411},
  number={1},
  pages={101--114},
  year={2016},
  publisher={Elsevier}
}

@article{Lauga2009,
	title={The hydrodynamics of swimming microorganisms},
	author={Lauga, Eric and Powers, Thomas R},
	journal={Reports on Progress in Physics},
	volume={72},
	number={9},
	pages={096601},
	year={2009},
	publisher={IOP Publishing}
}

@article{Blake1972,
	title={A model for the micro-structure in ciliated organisms},
	author={Blake, John},
	journal={Journal of Fluid Mechanics},
	volume={55},
	number={1},
	pages={1--23},
	year={1972},
	publisher={Cambridge University Press}
}

@article{Taylor1951,
  title={Analysis of the swimming of microscopic organisms},
  author={Taylor, Geoffrey I},
  journal={Proc. R. Soc. Lond. A},
  volume={209},
  number={1099},
  pages={447--461},
  year={1951}
}

@article{Blake1971,
  title={Infinite models for ciliary propulsion},
  author={Blake, JR},
  journal={Journal of Fluid Mechanics},
  volume={49},
  number={2},
  pages={209--222},
  year={1971},
  publisher={Cambridge University Press}
}

@article{Chrispell2013,
  title={An actuated elastic sheet interacting with passive and active structures in a viscoelastic fluid},
  author={Chrispell, John C and Fauci, Lisa J and Shelley, Michael},
  journal={Physics of Fluids},
  volume={25},
  number={1},
  pages={e1002167},
  year={2013},
  publisher={AIP}
}

@article{vogel2007living,
  title={Living in a physical world X. Pumping fluids through conduits},
  author={Vogel, Steven},
  journal={Journal of biosciences},
  volume={32},
  number={2},
  pages={207--222},
  year={2007},
  publisher={Springer}
}

@article{satir2007overview,
  title={Overview of structure and function of mammalian cilia},
  author={Satir, Peter and Christensen, S{\o}ren Tvorup},
  journal={Annu. Rev. Physiol.},
  volume={69},
  number={1},
  pages={377--400},
  year={2007},
  publisher={Annual Reviews}
}

@article{oyarte2025travelling,
  title={A travelling-wave strategy for plant--fungal trade},
  author={Oyarte Galvez, Loreto and Bisot, Corentin and Bourrianne, Philippe and Cargill, Rachael and Klein, Malin and van Son, Marije and van Krugten, Jaap and Caldas, Victor and Clerc, Thomas and Lin, Kai-Kai and others},
  journal={Nature},
  volume={639},
  number={8053},
  pages={172--180},
  year={2025},
  publisher={Nature Publishing Group UK London}
}

@article{chakrabarti2024cytoplasmic,
  title={Cytoplasmic stirring by active carpets},
  author={Chakrabarti, Brato and Rachh, Manas and Shvartsman, Stanislav Y and Shelley, Michael J},
  journal={Proceedings of the National Academy of Sciences},
  volume={121},
  number={30},
  pages={e2405114121},
  year={2024},
  publisher={National Academy of Sciences}
}

@article{jain2025geometric,
  title={Geometric effects in large scale intracellular flows},
  author={Jain, Olenka and Chakrabarti, Brato and Farhadifar, Reza and Gavis, Elizabeth R and Shelley, Michael J and Shvartsman, Stanislav Y},
  journal={PRX Life},
  volume={3},
  number={2},
  pages={023007},
  year={2025},
  publisher={APS}
}

@article{ul2022microscopic,
  title={Microscopic artificial cilia--a review},
  author={ul Islam, Tanveer and Wang, Ye and Aggarwal, Ishu and Cui, Zhiwei and Amirabadi, Hossein Eslami and Garg, Hemanshul and Kooi, Roel and Venkataramanachar, Bhavana B and Wang, Tongsheng and Zhang, Shuaizhong and others},
  journal={Lab on a Chip},
  volume={22},
  number={9},
  pages={1650--1679},
  year={2022},
  publisher={Royal Society of Chemistry}
}

@article{den2013microfluidic,
  title={Microfluidic manipulation with artificial/bioinspired cilia},
  author={den Toonder, Jaap MJ and Onck, Patrick R},
  journal={Trends in biotechnology},
  volume={31},
  number={2},
  pages={85--91},
  year={2013},
  publisher={Elsevier}
}

@article{vilfan2010self,
  title={Self-assembled artificial cilia},
  author={Vilfan, Mojca and Poto{\v{c}}nik, Anton and Kav{\v{c}}i{\v{c}}, Bla{\v{z}} and Osterman, Natan and Poberaj, Igor and Vilfan, Andrej and Babi{\v{c}}, Du{\v{s}}an},
  journal={Proceedings of the national academy of sciences},
  volume={107},
  number={5},
  pages={1844--1847},
  year={2010},
  publisher={National Academy of Sciences}
}

@article{gauger2009fluid,
  title={Fluid transport at low Reynolds number with magnetically actuated artificial cilia},
  author={Gauger, Erik M and Downton, Matthew T and Stark, Holger},
  journal={The European Physical Journal E},
  volume={28},
  number={2},
  pages={231--242},
  year={2009},
  publisher={Springer}
}

\newpage
\appendix

\begin{table}
\caption{\textbf{Biological ciliated channels:} mean ciliary fraction and confinement ratio estimated from published data.}
\label{tab:biological data}
\centering
\scriptsize
\setlength{\tabcolsep}{2.5pt}
\renewcommand{\arraystretch}{1.05}
\begin{tabularx}{\textwidth}{@{}l l l c c c l@{}}
\cline{1-7}
\multicolumn{7}{c}{\textbf{Ciliary carpets}} \\
\cline{1-7}
\textbf{Species} & \textbf{Common name} & \textbf{Organ} & \textbf{Coverage} & $\langle\Phi\rangle$ & $c$ & \textbf{References} \\
\cline{1-7}
\textit{Homo sapiens sapiens} & Human & Airway & 0.86 & 0.34 & 0.001 & \cite{roth2025structure}\\
\textit{Rattus norvegicus} & Rat & Brain ventricle & -- & 0.003 & 0.009 & \cite{kazanis2012number}\\
& & & & & & \cite{haemmerle2015neural}\\
\textit{Rattus norvegicus} & Rat & Airway & 0.53 & 0.21 & 0.01 & \cite{roth2025structure}\\
\textit{Mus musculus} & Mouse & Airway & 0.37 & 0.15 & 0.04 & \cite{Ramirez2020}\\
& & & & & & \cite{boucher2024high,li2021identification}\\
\textit{Homo sapiens sapiens} & Human & Fallopian tube & 0.34 & 0.14 & 0.20 & \cite{raidt2015ciliary}; \cite{varga2018lymphatic}\\
\textit{Danio rerio} & Larval zebrafish & Brain ventricle & -- & 0.02 & 0.22 & \cite{olstad2019ciliary}\\
\textit{Saccocirrus papillocercus} & Ringed worm & Foregut & 0.38 & 0.15 & 0.45 & \cite{purschke1996dorsolateral}\\
\textit{Dicyema acuticephalum} & Dicyema & Urn cavity & -- & 0.62 & 0.84 & \cite{furuya1997fine}\\
\cline{1-7}
\multicolumn{7}{c}{\textbf{Ciliary flames}} \\
\cline{1-7}
\textbf{Species} & \textbf{Common name} & \textbf{Organ} & \textbf{Coverage} & $\langle\Phi\rangle$ & $c$ & \textbf{References} \\
\cline{1-7}
\textit{Meiopriapulus fijiensis} & Penis worm & Protonephridia & -- & 0.59 & 0.61 & \cite{storch1989internal}\\
\textit{Thalassema thalassemum} & Spoon worm & Head kidney & -- & 0.73 & 0.67 & \cite{kato2011ultrastructure}\\
\textit{Artioposthia sp.} & Land planaria & Protonephridia & -- & 0.53 & 0.79 & \cite{rohde1992ultrastructure}\\
\textit{Habrotrocha rosa} & Rotifer & Protonephridia & -- & 0.92 & 0.83 & \cite{schramm1978excretory}\\
\textit{Lepidochitona corrugata} & Mollusk & Metanephridia & -- & 0.63 & 0.83 & \cite{Baeumler2012}\\
\textit{Urnatella gracilis} & Goblet worm & Protonephridia & -- & 0.88 & 0.85 & \cite{von1962erorterung}\\
\textit{Danio rerio} & Larval zebrafish & Metanephridia & -- & 0.80 & 0.87 & \cite{Ott2016}\\
\textit{Dugesia tigrina} & Planarian flatworm & Protonephridia & -- & 0.87 & 0.93 & \cite{Mckanna1968}\\
\textit{Aeolosoma bengalense} & Polychaete worm & Metanephridia & -- & 0.69 & 0.93 & \cite{bunke1994ultrastructure}\\
\textit{Geotrypetes seraphini} & Caecilian & Mesonephros & -- & 0.75 & 0.94 & \cite{mobjerg2004morphology}\\
\textit{Oikopleura dioica} & Larvacean & Ciliated funnel & -- & 0.90 & 1.00 & \cite{holmberg1982ciliated}; \cite{ling2024flow}\\
\cline{1-7}
\end{tabularx}
\end{table}

\newpage


\section{Estimating ciliary fraction and confinement ratio from data}
\label{App:biodata}
\begin{figure} [t]
    \centering
    \includegraphics[width=0.9\linewidth]{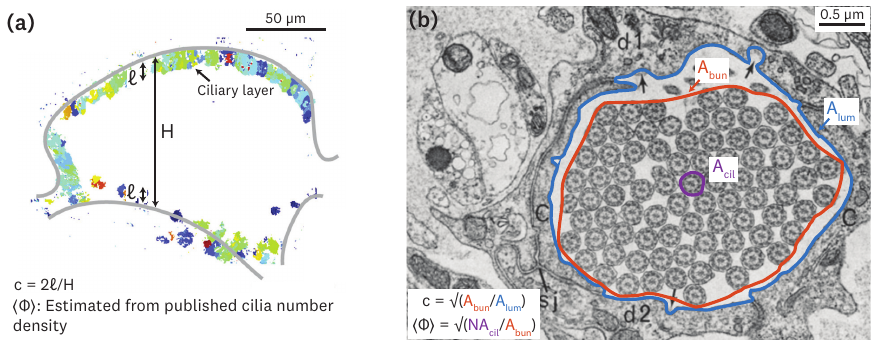}
    \caption{\textbf{Representative images of ciliary lumen from published literature.} (a) Cross section of the larval zebrafish brain ventricle \cite{olstad2019ciliary}. Colored regions indicate regions of ciliary activity. We measured the confinement ratio as the average of $\ell$ and $H$ taken from multiple locations throughout this cross section. The ciliary fraction was estimated from the published cilia number density \cite{olstad2019ciliary}. (b) Cross section of protonephridial ciliary flame in planaria \cite{Mckanna1968}. We manually traced the outlines of the lumen and the ciliary bundle to obtain their areas. The confinement ratio is computed as the square root of the bundle area divided by the lumen area. The ciliary fraction is the square root of the average area of one cilia multiplied by the number of cilia, divided by the bundle area.}
    \label{fig:bio data}
\end{figure}

We estimate from biological data two independent geometric quantities that enter the mathematical model: the mean ciliary fraction $\langle \Phi \rangle$ and the ciliary confinement ratio $c$. In our two-dimensional formulation, $\Phi(x,y,t)$ denotes the local ciliary fraction, defined as the area fraction occupied by ciliary material within the ciliated layer, and $\langle \Phi \rangle$ denotes its spatial average. The ciliary confinement ratio is a global measure of the fraction of the channel occupied by the ciliary layers, defined as $c = 2\ell/H$, where $\ell$ is the ciliary layer thickness and $H$ is the channel diameter.

The procedure used to estimate $\langle \Phi \rangle$ depends on the available biological data. When the cilia number density $\rho_c$ (number of cilia per unit area) is reported, we estimate the characteristic center-to-center spacing as $b = \rho_c^{-1/2}$ and define the corresponding mean ciliary fraction as $\langle \Phi \rangle \sim d\,\rho_c^{1/2}$. When $\rho_c$ is not available, we estimate it indirectly by combining a representative per-cell fraction (400 cilia per $100\,\mu\mathrm{m}^2$~\citep{nanjundappa2019regulation}) with the coverage fraction of ciliated cells, obtained from image analysis. The resulting $\rho_c$ is then used to compute $\langle \Phi \rangle$ as above.

For ciliary flames, the geometry differs from that of carpets, and cross-sectional images provide direct access to the relevant area fractions. We therefore estimate both $c$ and $\langle \Phi \rangle$ from images taken perpendicular to the channel axis. From these images, we trace the channel boundary, the extent of the ciliary bundle, and the individual cilia cross-sections. An example of this analysis is provided in figure \ref{fig:bio data}b. The ciliary layer thickness $2\ell$ is defined as the square root of the area occupied by the ciliary bundle and the lumen characteristic height $H$ is the square root of the lumen area. The confinement ratio $c$ is then defined as $c = 2\ell/H$, as above. The ciliary fraction $\langle \Phi \rangle$ is defined as the square root of the area occupied by cilia within the ciliary bundle, divided by the area of the bundle. This square root converts cross-sectional area fractions into effective linear fractions consistent with the reduced two-dimensional model (figure~\ref{fig:ref config and kin}). Refer to the supplementary data for notes on the analysis of each ciliated organ.

\begin{table} 
\centering
\algcaption{alg:precompute}{Precompute Eulerian Brinkman field $B(x,y,t^n)$ and ciliary velocity field $\bm{v}_c(x,y,t^n)$.}
\label{alg:inputs}
\hrule
\begin{minipage}{0.95\linewidth}
\begin{algorithmic}[1]
\State \textbf{Input:} $\bar{B}$, $\bm{\chi}_c(X,Y,t)$, current time $t^n$, Eulerian grid, time step $\Delta t$
\State Evaluate forward map: $\bm{\chi}_c^n(X,Y)=\bm{\chi}_c(X,Y,t^n)$
\State Compute deformation gradient: $\bm{F}^n=\nabla_{\bm{X}}\bm{\chi}_c^n$
\State Compute Jacobian: $J^n=\det\bm{F}^n$
\State Compute ciliary velocity: $\bm{v}_c^n(X,Y)=\partial_t\bm{\chi}_c(X,Y,t^n)$
\State Push forward Brinkman coefficient: $B^n(X,Y)=\bar{B}/J^n(X,Y)$
\For{each Eulerian grid point $(x,y)$}
\State Invert mapping: $(X,Y)=\bm{\chi}_c^{-1}(x,y,t^n)$
\State Set $\bm{v}_c^n(x,y)=\bm{v}_c^n(X,Y)$
\State Set $B^n(x,y)=B^n(X,Y)$
\EndFor
\State Store $\bm{v}_c^n(x,y)$ and $B^n(x,y)$
\end{algorithmic}
\end{minipage}
\end{table}


\section{Numerical flow solver}
\label{App:numerics}

To numerically solve the nondimensional Navier--Stokes--Brinkman equation \eqref{NSB} subject to the boundary conditions~\eqref{eq:bc}, we developed a custom pseudo-spectral solver in MATLAB. To this end, we expand the velocity field in Fourier modes in the periodic streamwise coordinate $x\in[0,1]$ and sine modes in the wall-normal coordinate scaled to the unit interval, $y/h\in[0,1]$, where $h=H/L$,
\begin{equation}
\bm{u}(x,y,t)
=
\sum_{m=-{M_x}/{2}}^{{M_x}/{2}-1}
\sum_{q=1}^{N_y}
\hat{\bm{u}}_{mq}(t)\,
e^{i\,2\pi m x}\,
\sin\!\left(\dfrac{\pi q y}{h}\right).
\end{equation}
Here, $M_x$ denotes the number of Fourier modes in the $x$ direction and $N_y$ the number of sine modes in the $y$ direction. We consider even $M_x$ and adopt the notation $m = -M_x/2, ..., M_x/2 -1$ accordingly. This choice of basis automatically enforces periodicity in $x$ and no-slip conditions at the walls $y=0$ and $y=h$. 

To rewrite~\eqref{NSB} on this $y$-scaled domain, we use the dimensionless gradient and Laplacian operators
\[
\nabla = \left(\partial_x, \dfrac{1}{h}\,\partial_y\right),
\qquad
\nabla^2 = \partial_{xx} + \dfrac{1}{h^2}\partial_{yy}.
\]
We solve the resulting equations numerically by employing a first-order implicit--explicit time-stepping scheme together with a projection method to enforce incompressibility. Specifically, we evaluate the nonlinear advection term $\bm{u}\cdot\nabla\bm{u}$ explicitly in physical space and the stiff linear terms (diffusion and Brinkman drag) implicitly in spectral space \citep{kutz2013data}, thereby avoiding the severe time-step restriction $\Delta t\lesssim \mathrm{Re}/B_{\max}$, where $B_{\max}$ denotes the largest value of the dimensionless Brinkman field. 

To this end, we write the nonlinear term in skew-symmetric form,
\begin{equation}
\label{eq:N}
    \bm{\mathcal{N}}(\bm{u}^n)
=\frac12\Big[(\bm{u}^n\cdot\nabla)\bm{u}^n
+\nabla\cdot(\bm{u}^n\otimes\bm{u}^n)\Big],
\end{equation}
and evaluate it in physical space. We substitute back into the Navier-Stokes-Brinkman~\eqref{NSB}, ignoring for now the incompressibility constraint, to obtain a set of linear equations for the intermediate velocity $\bm{u}^*$,
\begin{equation} \label{eq:NSB_intermediate}
\begin{aligned}
    \left(\mathcal{I}+\frac{\Delta t}{\mathrm{Re}}(-\nabla^2+B^{n+1}\mathbb{I})\right)\bm{u}^*
    =
    \bm{u}^n-\Delta t\,\bm{\mathcal N}(\bm{u}^n)
    +\frac{\Delta t}{\mathrm{Re}}\big(\Delta P\bm{e}_x+B^{n+1}\bm{v}_c^{n+1}\big)
\end{aligned}
\end{equation}
where $\mathcal{I}$ denotes the identity operator and $\mathbb{I}$ denotes the 2$\times$2 identity tensor. We solve this variable-coefficient linear system using a matrix-free Krylov method (MATLAB \verb|pcg|) with a diagonal preconditioner constructed from the constant reference field $\bar{B}$.

\begin{table}
\algcaption{alg:flow}{Implicit-explicit pseudo-spectral solver for the Navier--Stokes--Brinkman equations.}
\hrule
\begin{algorithmic}[1]
\Require Predefined adverse pressure gradient $\Delta P$, channel aspect ratio $h = H/L$
\Require Initial velocity $\bm{u}^0$, time step $\Delta t$
\For{$n=0,1,\dots,N_t-1$}
    \State Represent $\bm{u}^n(x,y)$ using Fourier modes in $x$ and sine modes in $y$
    \State Compute Eulerian ciliary fields $B(x,y,t^{n+1}$ and $\bm{v}_c(x,y,t^{n+1})$ according to Algorithm \ref{alg:precompute}
    \State Compute nonlinear term~\eqref{eq:N} in physical space
    \State Assemble linear system \eqref{eq:NSB_intermediate} in spectral space
    \State Solve \eqref{eq:NSB_intermediate} for intermediate velocity $\bm{u}^*$ using a matrix-free Krylov method and $\bar{B}$ preconditioner
    \State Solve the pressure Poisson equation \eqref{pressure poisson} in spectral space
    \State Project onto divergence-free velocity field \eqref{pressure increment}
    \State Store $\bm{u}^n(x,y)$
\EndFor
\State Compute and store streamwise $x$-averaged flow profile $u(y)$ and flow rate $Q(t)$ according to \eqref{eq:flow rate}
\end{algorithmic}
\end{table}

To enforce incompressibility, we project $\bm{u}^*$ onto a divergence-free field by seeking a pressure increment $\phi$ such that
\begin{equation} \label{pressure increment}
\bm{u}^{n+1}
=
\bm{u}^*
-
\frac{\Delta t}{\mathrm{Re}}\nabla\phi,
\qquad
\nabla\cdot\bm{u}^{n+1}=0.
\end{equation}
Taking the divergence of \eqref{pressure increment} yields the Poisson equation
\begin{equation} \label{pressure poisson}
\nabla^2\phi
=
\frac{\mathrm{Re}}{\Delta t}\,
\nabla\cdot\bm{u}^* .
\end{equation}
We solve equation~\eqref{pressure poisson} in spectral space using Fourier modes in $x$ and cosine modes in $y$, enforcing homogeneous Neumann boundary conditions at the walls. The divergence-free velocity $\bm{u}^{n+1}$ is then obtained from \eqref{pressure increment}. Algorithm~\ref{alg:flow} presents pseudocode for the numerical method, and 
full details of the validation and convergence of the scheme are provided in appendix~\ref{App:validate} and figures \ref{fig:convergence testing}-\ref{fig:k flow validation}.

\begin{figure} 
    \centering
    \includegraphics[width=\linewidth]{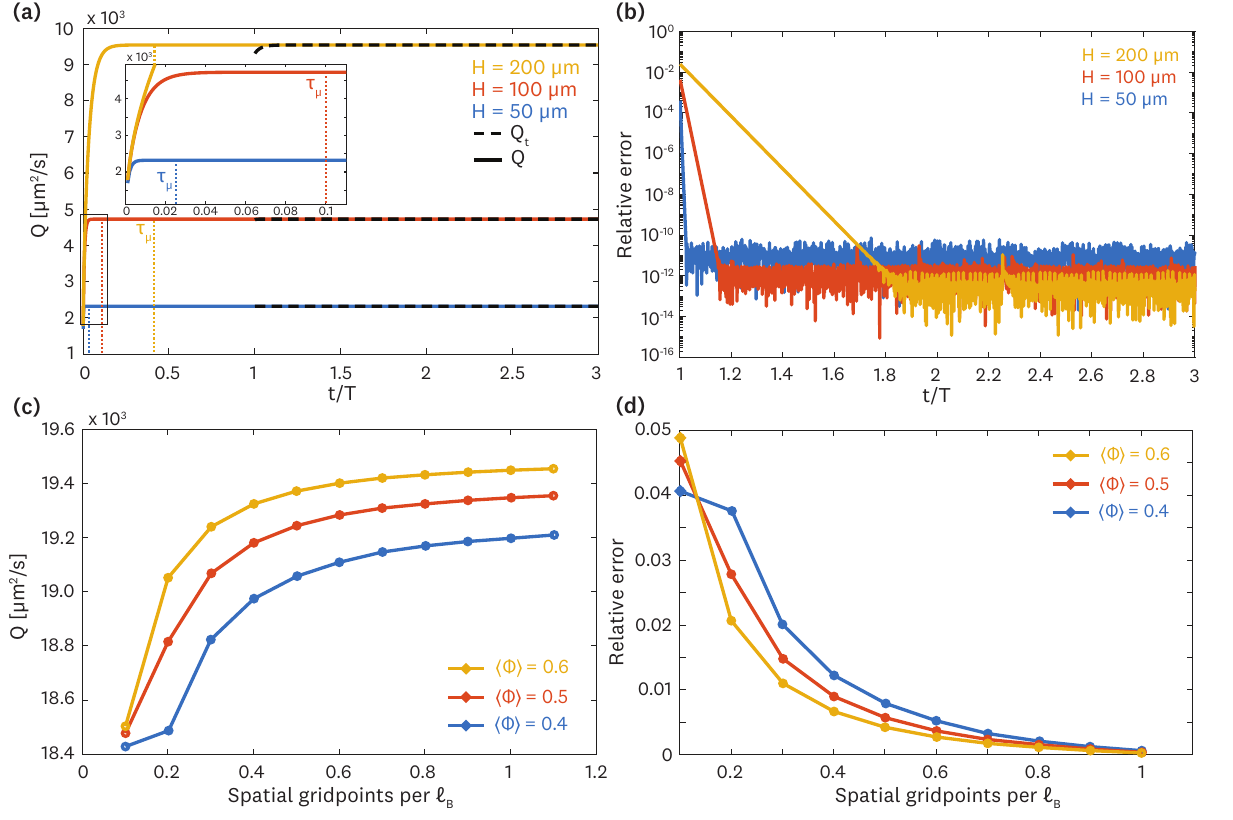}
    \caption{\textbf{Convergence of the numerical solver.}
    (a,b) Convergence to a periodic steady state. (c,d) Spatial resolution study.
    (a) Instantaneous flow rate obtained by spatial averaging over one metachronal wavelength (solid lines) is compared with the period-averaged flow rate computed from temporal averaging at a fixed location (dashed lines). Results are shown for $L=100\,\mu\mathrm{m}$ and $H=50,\,100,\,200\,\mu\mathrm{m}$ with $\Phi_{\rm ref}=0.2$, using one grid point per Brinkman screening length $\ell_B$ ($M_x=N_y=92,\,184,\,370$). The solution approaches a periodic steady state on the viscous diffusion timescale $\tau_\mu$, indicated by vertical dotted lines.
    (b) Relative difference between the two flow-rate measures. The comparison begins at $t/T=1$, once temporal averaging over a full beat period becomes available. The subsequent decay reflects convergence to the periodic steady state on the timescale $\tau_\mu$.
    (c) Steady-state flow rate $Q$ as a function of spatial resolution, measured in grid points per $\ell_B$, for $\Phi_{\rm ref}=0.4,\,0.5,\,0.6$.
    (d) Relative error in $Q$, computed with respect to the highest-resolution case (1.2 points per $\ell_B$). A minimum resolution of $M_x=N_y=32$ is imposed, which sets the lowest-resolution point.
    }
    \label{fig:convergence testing}
\end{figure}

We assess temporal convergence and determine the time required to reach a periodic steady state by comparing two measures of the flow rate. At each time step, we compute the spatially averaged flow rate $Q$ over one metachronal wavelength (defined in equation~\eqref{eq:flow rate}), and the period averaged flux at a given instant $t$
\begin{equation}
Q_{\rm t} = \frac{1}{T} \int_t^{t+T} \int_0^H u(x_0,y,t)\,dy\,dt.
\end{equation}
Agreement between $Q$ and $Q_{\rm t}$ indicates that a periodic steady state has been reached. We find that this occurs on the viscous diffusion timescale $\tau_\mu = \rho H^2/\mu$. In practice, we simulate for $1.25\,\tau_\mu$ and require that the relative change in $Q$ between successive time steps be below $5\times 10^{-4}$.

To determine spatial resolution, we vary the number of grid points relative to the Brinkman screening length $\ell_B$. We find that resolving $\ell_B$ with at least $0.4$ grid points yields relative errors below $O(10^{-2})$ in the flow rate. All simulations therefore use a minimum of $0.4$ grid points per $\ell_B$, with a resolution floor of $256\times256$ modes.

Temporal convergence is assessed by varying the time step over a range of resolutions. We observe first-order convergence, with relative errors in the flow rate below $O(10^{-3})$ at sufficiently small time steps. Based on this, we use $\Delta t = 2.5\times10^{-4}$, corresponding to 400 time steps per ciliary beat period, for all simulations in the main text.

\section{Validation of numerical flow solver}
\label{App:validate}

\begin{figure}
    \centering
    \includegraphics[width=\linewidth]{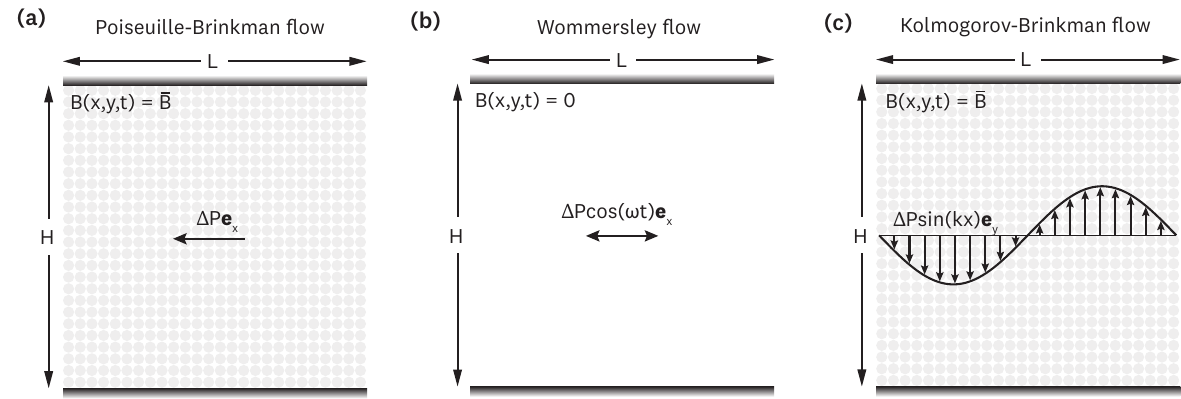}
    \caption{\textbf{Validation tests.}
    (a) Poiseuille–Brinkman flow driven by a constant body force $\Delta P\,\bm{e}_x$ in a homogeneous porous medium with Brinkman coefficient $\bar{B}$.
    (b) Womersley flow driven by an oscillatory body force $\Delta P \cos(\omega t)\,\bm{e}_x$ in the absence of Brinkman drag ($\bar{B}=0$).
    (c) Kolmogorov–Brinkman flow driven by a spatial body force $\Delta P\sin(kx)\,\bm{e}_y$ with homogeneous Brinkman coefficient $\bar{B}$.}
    %
    \label{fig:validation intro}
\end{figure}

To validate the pseudo-spectral Navier-Stokes–Brinkman solver and the projection scheme, we compare numerical solutions against analytical results for three canonical flows: (i) pressure-driven Poiseuille–Brinkman flow, (ii) oscillatory Womersley flow, and (iii)  wall-bounded Kolmogorov flow with homogeneous Brinkman drag illustrated in figure \ref{fig:validation intro}. In each case, the numerical solution is advanced in time until a steady or periodic steady state is reached and is then compared to the corresponding closed-form solution.

We first consider the classical Poiseuille flow corresponding to flow between two no-slip walls in the absence of a porous medium.  The nondimensional Navier--Stokes equations~\eqref{NSB} reduces to the well-known one-dimensional equation, subject to no-slip boundary conditions at the channel walls,
\begin{equation}
\frac{d^2 u(y)}{dy^2} = -\,\Delta P , \qquad u(0)=u(H)=0.
\end{equation}
Integrating twice yields the classical parabolic profile,
\begin{equation}
u(y)
=
\frac{\Delta P}{2}\,y\,(H-y).
\label{eq:Poiseuille}
\end{equation}
We next consider the Poiseuille--Brinkman flow, by including a homogeneous, static Brinkman medium by setting $\bm{v}_c=\bm{0}$ and $B = \bar{B}$. The steady governing equation, again with no-slip boundary conditions, becomes
\begin{equation}
\frac{d^2 u(y)}{dy^2} - \bar{B}\,u(y) = -\,\Delta P , \qquad u(0)=u(H)=0,
\end{equation}
for which
the analytical solution is given by
\begin{equation}
u(y)
=
\frac{\Delta P}{\bar{B}}
\left[
1
-
\frac{\cosh\!\left(\sqrt{\bar{B}}\left(y-\tfrac{H}{2}\right)\right)}
     {\cosh\!\left(\tfrac{\sqrt{\bar{B}} H}{2}\right)}
\right].
\label{eq:PoiseuilleBrinkman}
\end{equation}

\begin{figure} [t]
    \centering 
    \includegraphics[width=1\linewidth]{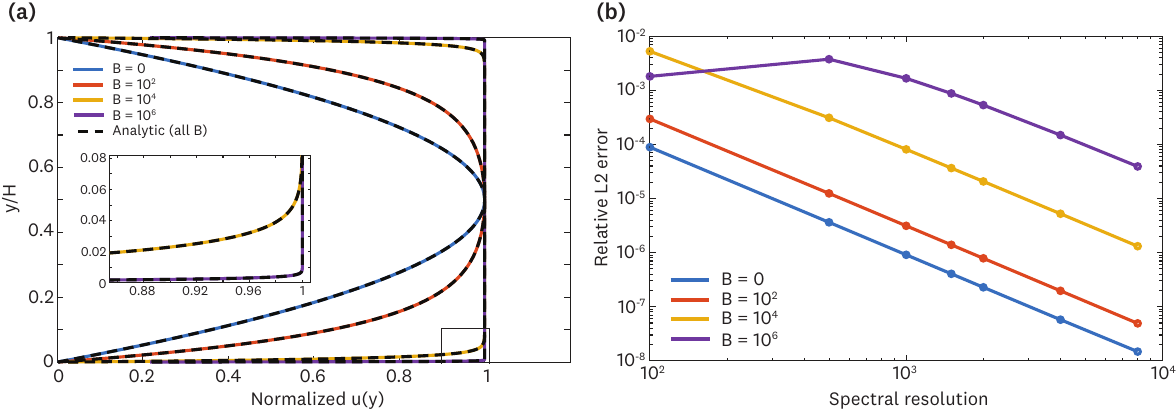}
    \caption{\textbf{Validation using Poiseuille–Brinkman flow.}
    (a) Normalized velocity profiles for pressure-driven flow with homogeneous Brinkman coefficient $B=0$--$10^{6}$ at the highest tested resolution $M_x = N_y = 8000$. Numerical solutions (solid lines) agree with analytical solutions (dashed lines). Each profile is normalized by its maximum; increasing $B$ sharpens the near-wall gradients (inset).
    (b) Relative $L^2$ error versus spectral resolution $M_x=N_y$. Second-order spatial convergence is observed once the Brinkman screening length is resolved. Parameters: $H=L=100\,\mu\mathrm{m}$, $\mathrm{Re}=0.01$, $\Delta P=0.1$.
    } 
    \label{fig:pb flow val}
\end{figure}
Testing these cases in the numerical solver corresponds to setting  $\bar{B}=0$ and $\bm{v}_c=\bm{0}$ for the Poiseuille flow, and setting $B=\bar{B}$ everywhere with $\bm{v}_c=\bm{0}$ for the Poiseuille--Brinkman flow, while imposing a constant adverse pressure gradient $\Delta P$. The solution is advanced in time to steady state and compared with the analytical profiles~\eqref{eq:Poiseuille} and~\eqref{eq:PoiseuilleBrinkman}. As shown in figure~\ref{fig:pb flow val}, excellent agreement is obtained, confirming second-order spatial convergence and validating the numerical treatment of the viscous and Brinkman terms.

\begin{figure} 
    \centering
    \includegraphics[width=\linewidth]{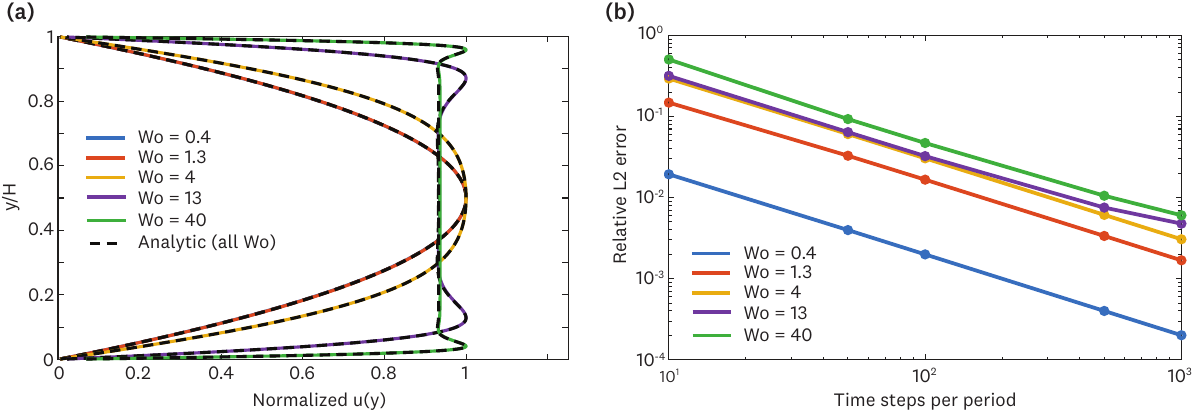}
    \caption{\textbf{Validation using oscillatory Womersley flow.}
    (a) Normalized maximum amplitude velocity profiles for Womersley numbers $\mathrm{W}_{\rm o}=0.4$--$40$ at high temporal (1000 time steps per period) and spatial ($M_x=N_y=800$) resolution. Numerical (solid) and analytical (dashed) solutions are indistinguishable; the profiles for $\mathrm{W}_{\rm o}=0.4$ and $1.3$ overlap.
    (b) Relative $L^2$ error, averaged over one period at steady state, versus time steps per period. First-order temporal convergence is observed. Parameters: $\omega = [0.1,1,10,100,1000] \ 
    \rm{Hz}$, $H=L=1\ \rm{mm}$, $\rm{Re}=1, B=0$, $\Delta P=0.01$.
    }
    \label{fig:wom flow}
\end{figure}

To validate the temporal integration scheme and the implicit treatment of viscous diffusion, we consider oscillatory pressure-driven flow between two no-slip walls (Womersley flow). In the absence of Brinkman drag and ciliary forcing, the equations reduce to a one-dimensional unsteady Stokes problem with no-slip boundary conditions,
\begin{equation}
\frac{\partial u}{\partial t}
=
\frac{1}{\mathrm{Re}}\frac{\partial^2 u}{\partial y^2}
+
\Delta P \cos(\omega t),
\qquad
u(0,t)=u(H,t)=0,
\label{eq:womersley_pde}
\end{equation}
where $u(y,t)$ is the streamwise velocity and $\Delta P\cos(\omega t)$ is a spatially uniform, oscillatory body force representing an imposed oscillatory pressure gradient. We seek a time-harmonic analytical solution of the form $u(y,t)=\Re\!\left\{\hat{u}(y)\,e^{i\omega t}\right\}$,
where $\Re\{\cdot\}$ denotes the real part of a complex quantity. Substitution into \eqref{eq:womersley_pde} yields a second-order boundary-value problem for the complex amplitude $\hat{u}(y)$. Solving for $\hat{u}(y)$, multiplying by $e^{i\omega t}$, and taking the real part we arrive at
\begin{equation}
u(y,t)=\Re\!\left\{\frac{\Delta P}{i\omega}
\left[1-
\frac{\cosh\!\left(\sqrt{i\omega \mathrm{Re}}\left(y-\tfrac{H}{2}\right)\right)}
     {\cosh\!\left(\tfrac{\sqrt{i\omega \mathrm{Re}}\,H}{2}\right)}
\right]
e^{i\omega t}
\right\}.
\label{eq:womersley_solution}
\end{equation}
Figure~\ref{fig:wom flow}a shows normalized analytical and numerical velocity profiles at peak amplitude for the forcing frequencies considered. The Womersley number, $\mathrm{Wo} = (H/2)\sqrt{\rho \omega / \mu}$, measures the ratio of oscillatory to viscous timescales. Excellent agreement is obtained at sufficient temporal resolution. Varying the time step confirms first-order temporal convergence (figure~\ref{fig:wom flow}b).

Lastly, to validate the full spectral solver and projection method, we consider steady, wall-bounded Kolmogorov flow with homogeneous Brinkman drag. The governing equations are
\begin{equation} \label{k flow}
\begin{aligned}
-\nabla p + \nabla^2 \bm{u} - \bar{B}\,\bm{u} + \Delta P \sin(kx)\,\bm{e}_y = 0, 
\qquad \nabla \cdot \bm{u} = 0,
\end{aligned}
\end{equation}
subject to periodic boundary conditions in $x$ and no-slip conditions at $y=0$ and $y=H$. The body force $\Delta P \sin(kx)\bm{e}_y$ varies sinusoidally in $x$ and drives a fully two-dimensional flow.

\begin{figure} [t!]
    \centering
    \includegraphics[width=1\linewidth]{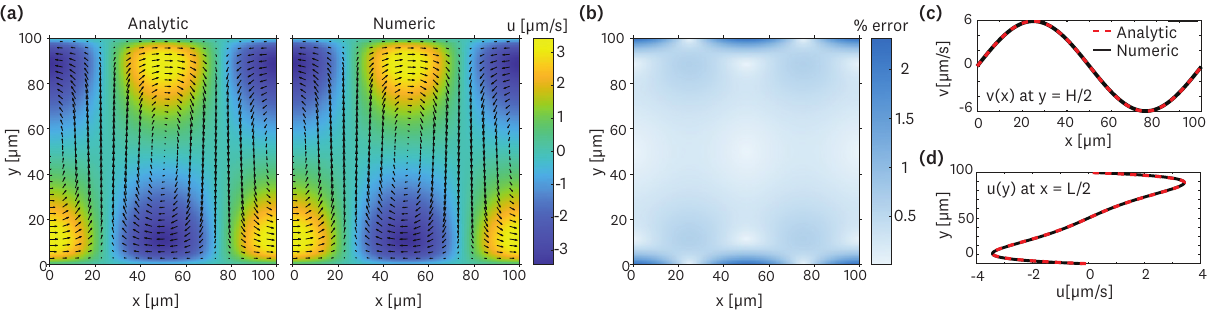}
    \caption{\textbf{Validation using wall-bounded Kolmogorov-Brinkman flow.} (a) Analytical flow field (left) and numerical flow field (right). Color bar indicates the magnitude of the horizontal component of the flow $u$. (b) Percent error between the analytical and numerical flow fields, as measured by $100 \times |\bm{u}_{num}-\bm{u}_{ana}|/(\max(|\bm{u}_{ana}|)).$ (c) Vertical component of the flow $v(x)$ at $y = H/2$. (d) Horizontal component of the flow $u(y)$ at $x = L/2$. Red dashed line indicates the analytical curve, solid black the numerical curve. Here, $M_x = 480$ and $N_y = 960$. The wavenumber $k$ is set so that there's a single waveform per $L$. Parameter values are $L = H = 100 \ \rm \mu m$, $\rm Re = 1$, $\bar{B} = 100$, and $\Delta P = 10^{-3}$.}
    \label{fig:k flow validation}
\end{figure}

We seek a single Fourier-mode solution of the form
\begin{equation}
u(x,y)=U(y)\cos(kx), \qquad
v(x,y)=V(y)\sin(kx), \qquad
p(x,y)=P(y)\sin(kx).
\end{equation}
Substitution into~\eqref{k flow} and incompressibility yield
\begin{equation}
V'(y)=k\,U(y),
\end{equation}
which allows elimination of $V$ and $P$, leading to the fourth-order ordinary differential equation
\begin{equation}
U''''-(2k^2+\bar{B})U''+(k^4+\bar{B} k^2)U=0.
\label{eq:4thODE}
\end{equation}
Let $\lambda = \sqrt{k^2+\bar{B}}$. The general solution of~\eqref{eq:4thODE} is
\begin{equation}
U(y)=
A\sinh(ky)+ \beta \cosh(ky)
+ C\sinh(\lambda y)+D\cosh(\lambda y).
\end{equation}
The constants are determined from the no-slip conditions $U(0)=U(H)=0$ and the wall momentum balance
\begin{equation}
U'''(y)-(2k^2+\bar{B})U'(y)=k\Delta P
\quad \text{at} \quad y=0,H.
\end{equation}
Solving for $U(y)$ first, then  for $V(y)$ and $P(y)$, we get
\begin{equation}
\begin{aligned}
U(y)
&= A\sinh(ky) + \beta\cosh(ky)
 + C\sinh(\lambda y) + D\cosh(\lambda y), \\[4pt]
V(y)
&= A\bigl[\cosh(ky)-1\bigr] + \beta\sinh(ky)
 + \frac{k}{\lambda}C\bigl[\cosh(\lambda y)-1\bigr]
 + \frac{k}{\lambda}D\sinh(\lambda y), \\[4pt]
P(y)
&= -\frac{\bar{B}}{k}\Bigl(A\sinh(ky)+\beta\cosh(ky)\Bigr).
\end{aligned}
\end{equation}
The coefficients satisfy
\begin{equation}
\begin{aligned}
A = -\frac{\Delta P}{\lambda^2}-\frac{k}{\lambda}C, \qquad
\beta = -D,
\end{aligned}
\end{equation}
with $C$ and $D$ given by
\begin{equation}
\begin{aligned}
C
&= \Delta P\,
\frac{ \frac{k}{\lambda}\sinh(kH)\bigl(\cosh(\lambda H)-\cosh(kH)\bigr)
 -(\cosh(kH)-1)\Bigl(\sinh(\lambda H)-\frac{k}{\lambda}\sinh(kH)\Bigr)
  }
  { k\lambda\bigl(\cosh(\lambda H)-\cosh(kH)\bigr)^2
 +\Bigl(\sinh(\lambda H)-\frac{k}{\lambda}\sinh(kH)\Bigr)
\bigl(\lambda^2\sinh(kH)-k\lambda\sinh(\lambda H)\bigr)}, \\[8pt]
D
&= \Delta P\,
\frac{ (\cosh(\lambda H)-\cosh(kH))
 \Bigl(\cosh(kH)-1-\frac{k}{\lambda}\sinh(kH)\Bigr)
  }
  { k\lambda\bigl(\cosh(\lambda H)-\cosh(kH)\bigr)^2
 +\Bigl(\sinh(\lambda H)-\frac{k}{\lambda}\sinh(kH)\Bigr)
\bigl(\lambda^2\sinh(kH)-k\lambda\sinh(\lambda H)\bigr)}.
\end{aligned}
\end{equation}
To simulate this flow numerically, we impose the sinusoidal body force in~\eqref{k flow} and advance the solution in time to steady state, defined by a relative change in the maximum velocity below $10^{-4}$ between successive time steps. Figure~\ref{fig:k flow validation} shows excellent agreement between numerical and analytical solutions over all parameter values tested, validating the solver for fully two-dimensional flows.

In all validation simulations, the maximum divergence remained $O(10^{-12})$ or smaller, confirming that the projection method (appendix~\ref{App:numerics}) enforces incompressibility to machine precision.

\end{document}